\documentclass[aps,pra,twocolumn,showpacs,superscriptaddress]{revtex4}
\usepackage{amsthm}
\usepackage{newlfont}
\usepackage{graphicx}
\usepackage{subfigure}
\usepackage[section]{placeins}
\usepackage{psfrag}
\usepackage{color}
\usepackage{caption2}
\usepackage{flafter}
\usepackage{bm}
\usepackage{amsthm}
\usepackage{newlfont}
\usepackage{graphicx}
\usepackage{subeqn}
\usepackage{subfigure}
\usepackage[section]{placeins}
\usepackage{psfrag}
\usepackage{caption2}
\usepackage{flafter}
\usepackage{bm}
\usepackage{bbding}
\usepackage{pifont}
\usepackage{wasysym}
\usepackage{amssymb}
\usepackage{dcolumn}
\begin{document}
\title{Modulational instability in non-Kerr photonic Lieb lattice with metamaterials}
\author{A. K. Shafeeque Ali}
\affiliation{\normalsize \noindent Centre for Nonlinear Dynamics, School of Physics,
Bharathidasan University, Tiruchirappalli 620 024, India.}
\author{Andrei I. Maimistov}
\affiliation{\normalsize \noindent Department of Solid State Physics
and Nanostructures, National
Nuclear Research University,\\
Moscow Engineering Physics Institute, Moscow 115 409.}
\author{K. Porsezian}
\affiliation{\normalsize \noindent Department of Physics, Pondicherry University, Pondicherry
  605 014, India.}
\author{A. Govindarajan}
\affiliation{\normalsize \noindent Centre for Nonlinear Dynamics, School of Physics,
Bharathidasan University, Tiruchirappalli 620 024, India.}
\author{M. Lakshmanan}
\affiliation{\normalsize \noindent Centre for Nonlinear Dynamics, School of Physics,
Bharathidasan University, Tiruchirappalli 620 024, India.}
\date{\today}
\begin{abstract}
We present an analysis of modulational instability of diffractionless waves in a face-centered square lattice of waveguides featuring non-Kerr nonlinearity, which are constituted
by a combination of positive and negative refractive indices. The unit cell of the lattice
 consists of three different waveguides with different
optical properties. The dispersion curve of the lattice supports flat bands and thereby the base equations describing the model have
particular solutions that correspond to the diffractionless waves
propagating along the waveguides.  We also observe a unique ramification of nonlinearities in controlling the flat bands optically. The diffractionless wave solutions are derived and the stability of these
distributions are investigated in a nutshell by adopting the standard linear stability approach.
\end{abstract}


\maketitle
\section{Introduction}
In recent studies pertaining to optical lattices, there has been a surge of interest in the investigation of photonic spectrum with flat band \cite{Flach:Leykam:14,Longi:14,Maimis:15,G:Malomed:16,Daniel,Leykam1}. In
\cite{Mukherjee:15,Mukherjee:Spracklen:15a} the effect of flat band
was investigated experimentally. As is well-known, the knowledge about the existence of flat band states in photonic lattices has accelerated quantum simulation of flat band models in a highly controllable environment \cite{Mukherjee:Spracklen:15a}. The appearance of flat band means that
for some particular electromagnetic field distributions inside the waveguide
 discrete diffraction is absent. The diffractionless propagation
of electromagnetic waves in waveguide arrays are discussed in
\cite{Vicencio:14,Vicencio:15,Fang:15,Maim:Gabi:16,Maim:16,Mukherjee2,Bastian}.
 The conditions to observe the propagation of diffractionless modes in the presence of Kerr nonlinearity in the Lieb lattice have been investigated \cite{ppbelicev}.
Further, non-Hermiticity-induced flat band in a parity-time symmetric photonic lattice and controllable localization of light in the non-Hermitian systems as a result of flat band have been reported \cite{Hamidreza}. Distortion-free image transmission in a two-dimensional perovskite-like photonic structure as a result of superposition of localized flat-band states has been demonstrated \cite{Shiqiang}. Also, high-fidelity transmission of the complex patterns in a two-dimensional pyrochlore-like photonic structure due to the linear superposition of the flat band eigen modes of the Kagome lattices has been verified \cite{Yuanyuan}. The bifurcation of families of localized discrete solitons from the localized linear modes of the flat band with zero power threshold in a two dimensional Kagome lattice with defocusing nonlinearity has also been studied \cite{vice}.  The stability of the flat band solution in a  rhombic nonlinear optical waveguide array breaks down when the intensity per waveguide exceeds the threshold value \cite{maimistov17}.
\par
On the other hand, theoretical investigation on the nonlinear pulse propagation in waveguide arrays has been receiving considerable attention. For instance, observation of discrete spatial optical solitons in an array has been reported \cite{Eisenberg}. When the initial excitation is not centered on a waveguide, the discrete solitons in a waveguide array can acquire transverse momentum and propagate at an angle with respect to the waveguide direction \cite{Morandotti}. In the year 2010, the first experimental observation of three-dimensional light bullets in waveguide arrays featuring quasi-instantaneous cubic nonlinearity and a periodic, transversally modulated refractive index has been reported \cite{Minardi}. The space time coupling in a waveguide array breaks the spectral symmetry of light bullets to a considerable degree and modifies their group velocity,
leading to superluminal propagation when the light bullets decay \cite{Falk}. Also, bright and dark spatial gap solitons have been demonstrated in waveguide arrays \cite{Mandelik}. Dispersive shock waves in the nonlinear waveguide arrays have also been studied experimentally \cite{Shu}.
 \par
 Besides the above, study on waveguide arrays with positive and negative refractive index waveguides have also received considerable attention due to their unique features. Finite gap solitons observed in the arrays with positive and negative refractive index waveguides having nonlinearities of different types show the phenomenon of  symmetry breaking in the Fourier space \cite{Zezyulin}. Both staggered and unstaggered discrete solitons formed in positive and negative refractive index material waveguide arrays can become highly localized states near the zero diffraction points even for low powers \cite{Alexander}. The interaction of the nonlinear effects of the channels has great influence on the generation of the modulation instability \cite{Lingling}. The modulation instability of condensate solution for electromagnetic wave propagating in such waveguide arrays  has also been investigated  recently \cite{shaf}.
  \par
  \begin{figure}[t]
    \center{\includegraphics[scale=0.75]{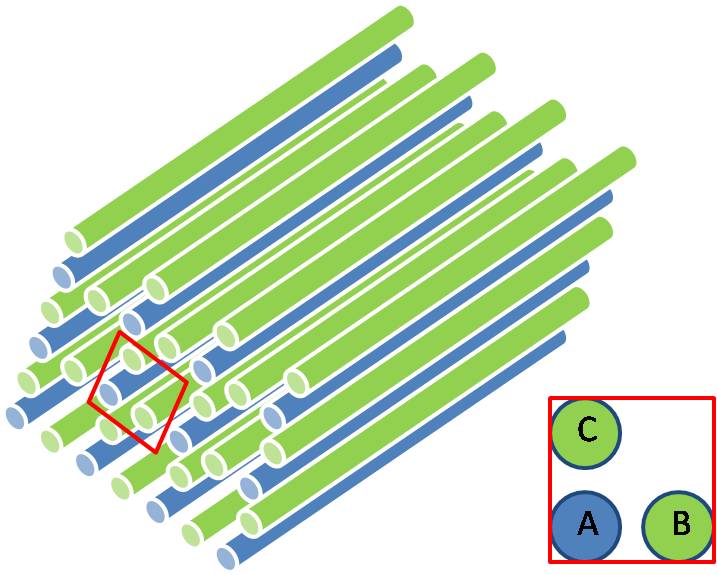}}
     \caption{Schematic diagram of a two dimensional waveguide array with alternating sign of refractive index(left) and a unit cell with waveguides A, B and C of different
optical properties(right). }
     \label{Fig:1}
\end{figure}
In this paper, we study the propagation of coupled electromagnetic waves in a two dimensional
waveguide array, which consists of waveguides with positive and negative refractive indices. The
cross section of the array is in the form of face centered square lattice. All waveguides are
assumed to be nonlinear with cubic, quintic and septimal nonlinearities:
\begin{equation}
P_{nl}=\chi^{(3)}|E|^2-\chi^{(5)}|E|^4+\chi^{(7)}|E|^6,
\end{equation}
where $\chi^{(n)}$ is the $n$-th order nonlinear susceptibility  and $E$ is the electric field strength
of the wave connected to the waveguide.
\par
The structure of two dimensional waveguide array  considered in this study can be realized by arranging unit cells of the array in periodic manner as shown in Fig. \ref{Fig:1}. Each unit cell consists three waveguides  of different optical properties. The attainment of modern technologies, such as nanotechnology, enables the manufacturing of such waveguide arrays  with
unusual and different optical properties including negative refraction \cite{smith1,shelby,shelby2}. The phenomenon of negative refraction can be applied in different optical components for integrated and fiber optics \cite{scot,jason}. The nonlinear properties of metamaterials can be obtained using nonlinear insertions, an element showing nonlinear response  such as diodes to resonant meta atoms \cite{lapine}. Nonlinear insertions are suitable to obtain high nonlinear response with a few watts of
power at microwave and lower terahertz frequencies, but
this method fails to develop nonlinear NIM at optical frequencies.  At optical range, desired nonlinear response can be obtained by embedding the metaatoms into a nonlinear dielectric medium \cite{agran}. Also, the lossless metamatrials at optical range can be realized by the implantation of components with active molecules into the structure of artificial materials \cite{Shumin}. Moreover, metamaterial permits engineering of material parameters from their basic constituents \cite{Shad,Tass}. This characteristics provide the tuning of material parameters at will.
\par
We will theoretically study the existence of flat bands in such a nonlinear waveguide array and hence focus on the investigation of diffractionless solution of the system of equations describing the evolution of the electromagnetic fields. The crucial instability of such diffractionless solution is also investigated using modulational instability analysis. It is found that the photonic band structure as well as the stability of flat band modes are  highly dependent on coefficient functions $\kappa(k)$ and higher order nonlinearities.  Hence, the stable propagation of electromagnetic waves can be achieved by tuning these parameters in the lattice. Also, this study suggests the possibility for optically controlling the band structures of waveguide arrays.
\par
The paper is organized as follows. Following a self contained introduction, in Sec. II the theoretical model and solutions of the problem  are presented. The diffractionless solutions and their stability are studied in Sec. III. In Sec. IV investigation on modulational instability in metamaterial waveguide arrays is carried out in detail followed by a short presentation of the summary and conclusion in Sec. V.
\section{Theoretical Model and Solutions}
 Consider the waveguide array arranged in a plane as shown in Fig. \ref{Fig:1}. The cross section of this configuration has face-centered square lattice structure with alternating signs of refractive index. The unit cell of the structure consists of three nonlinear waveguides with cubic, quintic and septimal nonlinearities. The system of equations describing the evolution of the envelopes of the wave localized in
the waveguide of the unit cell can be derived by using the procedure developed in \cite{shaf} by adopting the tight-binding approximation. The resulting system of equations is as follows,
\begin{subequations}
 \label{eq:ABC:gener:1}
\begin{eqnarray}
 i(\frac{\partial }{\partial \tau}+ \sigma_1\frac{\partial}{\partial \zeta}) A_{n,m} + \alpha_1 (B_{n,m}+B_{n-1,m})
e^{i\delta_{ba}\zeta}  \nonumber \\
 + \alpha_2(C_{n,m}+C_{n,m-1})e^{i\delta_{ca}\zeta} +r_{11}|A_{n,m}|^2A_{n,m}
\nonumber \\-r_{12}|A_{n,m}|^4A_{n,m}+r_{13}|A_{n,m}|^6A_{n,m} =0,\label{eq:A:1}
\end{eqnarray}
\begin{eqnarray}
 i\left(\frac{\partial }{\partial \tau}+ \sigma_2\frac{\partial}{\partial \zeta} \right) B_{n,m} + \alpha_1 (A_{n,m}+A_{n+1,m})
e^{-i\delta_{ba}\zeta} \nonumber \\
 \qquad\qquad + r_{21}|B_{n,m}|^2B_{n,m}-r_{22}|B_{n,m}|^4B_{n,m} \nonumber \\+r_{23}|B_{n,m}|^6B_{n,m}=0,\label{eq:B:1}
 \end{eqnarray}
 \begin{eqnarray}
 i\left(\frac{\partial }{\partial \tau}+ \sigma_3\frac{\partial}{\partial \zeta} \right) C_{n,m} + \alpha_2 (A_{n,m}+A_{n,m+1})
e^{-i\delta_{ca}\zeta} \nonumber \\
+ r_{31}|C_{n,m}|^2C_{n,m} -r_{32}|C_{n,m}|^4C_{n,m}\nonumber \\+r_{33}|C_{n,m}|^6C_{n,m}  =0.\label{eq:C:1}
\end{eqnarray}
\end{subequations}
Here $\sigma_j$ represents the sign of refractive index of individual waveguides. Hence, for a positive refractive index waveguide $\sigma_j=1$,
and for a negative refractive index waveguide $\sigma_j=-1$. The
pair  $(n, m)$ of integers stands for the unit cell label in the two dimensional lattice.
Also $A_{n,m}$, $B_{n,m}$ and $C_{n,m}$ are the normalized
envelopes of the associated wave localized in the appropriate waveguide of the unit cell
with indices $(n,m)$. Further, $\delta_{ba} = \beta_b-\beta_a $ and
$\delta_{ca} = \beta_c-\beta_a $ are the mismatch between the
wave numbers (propagation constants) $\beta_a$, $\beta_b$ and $\beta_c$. The parameters $\alpha_1$
and $\alpha_2$ indicate the coupling strengths
between neighboring waveguides. Also, $r_{11}$, $r_{21}$ and $r_{31}$ are cubic nonlinear coefficients, $r_{12}$, $r_{22}$ and $r_{32}$ are quintic nonlinear coefficients and $r_{13}$, $r_{23}$ and $r_{33}$ are septimal nonlinear coefficients.
\par
 Now, let us choose for convenience the various nonlinear coefficients as $r_{11}=r_{21}=r_{31}=R_1$,
 $r_{12}=r_{22}=r_{32}=R_2$ and $r_{13}=r_{23}=r_{33}=R_3$. Also, we introduce the following transformations of the fields,
\begin{eqnarray}
  && A_{n, m} = \tilde{A}_{n, m}e^{i\delta_{ba}\zeta/2}, \quad B_{n, m} =
\tilde{B}_{n, m}e^{-i\delta_{ba}\zeta/2},  \nonumber \\
 && \qquad \quad C_{n, m} = \tilde{C}_{n,
m}e^{i(\delta_{ba}/2-\delta_{ca}) \zeta},\nonumber
\end{eqnarray}
so that the system of Eqs. (\ref{eq:ABC:gener:1}) takes the following form,
\begin{subequations}
\begin{eqnarray}
 i\left(\frac{\partial}{\partial \tau}+\sigma_1 \frac{\partial}{\partial \zeta}\right)\tilde{A}_{n, m}
  -\frac{1}{2}\sigma_1\delta_{ba}\tilde{A}_{n, m}\nonumber\\
   +\alpha_1 (\tilde{B}_{n, m}+\tilde{B}_{n-1,m})
   +\alpha_2 (\tilde{C}_{n, m}+\tilde{C}_{n, m-1})\nonumber\\
   + R_{1}|\tilde{A}_{n, m}|^2\tilde{A}_{n, m} -R_{2}|\tilde{A}_{n, m}|^4\tilde{A}_{n, m}\nonumber\\+R_{3}|\tilde{A}_{n, m}|^6\tilde{A}_{n, m}=0,
\end{eqnarray}
\begin{eqnarray}
 i\left(\frac{\partial}{\partial \tau}+\sigma_2 \frac{\partial}{\partial \zeta}\right)\tilde{B}_{n, m}
  + \frac{1}{2}\sigma_2\delta_{ba} \tilde{B}_{n, m} \nonumber \\
  +\alpha_1 (\tilde{A}_{n, m}+\tilde{A}_{n+1, m}) +R_{1}|\tilde{B}_{n, m}|^2\tilde{B}_{n, m}\nonumber \\-R_{2}|\tilde{B}_{n, m}|^4\tilde{B}_{n, m}+R_{3}|\tilde{B}_{n, m}|^6\tilde{B}_{n, m}=0,
  \end{eqnarray}
  \begin{eqnarray}
  i\left(\frac{\partial}{\partial \tau}+\sigma_3 \frac{\partial}{\partial \zeta}\right)\tilde{C}_{n, m}
  -\sigma_3\varphi_0 \tilde{C}_{n, m} + \nonumber \\
  \qquad \qquad + \alpha_2 (\tilde{A}_{n, m}+\tilde{A}_{n,
  m+1}) +R_{1}|\tilde{C}_{n, m}|^2\tilde{C}_{n, m}\nonumber \\-R_{2}|\tilde{C}_{n, m}|^4\tilde{C}_{n, m}+R_{3}|\tilde{C}_{n, m}|^6\tilde{C}_{n, m}=0,
\end{eqnarray}
\end{subequations}
where $\varphi_0=\delta_{ba}/2-\delta_{ca} $.  Let us assume the waveguides B and C are identical, admitting the same propagation constants ($\beta_b =\beta_c$),
  so that $\delta_{ba}/2=\delta_{ca}/2 = \Delta$, where
$\Delta=-\varphi_0$. Also, let us consider
the case of $\sigma_2=\sigma_3\equiv\sigma=\pm 1$ and $\sigma_1=1$. Thus
the new system of equations reads as
\begin{subequations}
\begin{eqnarray}
   i\left(\frac{\partial}{\partial \tau}+ \frac{\partial}{\partial \zeta}\right)\tilde{A}_{n, m}
  -\Delta\tilde{A}_{n, m}\nonumber\\
    +\alpha_1 (\tilde{B}_{n, m}+\tilde{B}_{n-1,m}) \nonumber\\
  +\alpha_2 (\tilde{C}_{n, m}+\tilde{C}_{n, m-1})
   + R_{1}|\tilde{A}_{n, m}|^2\tilde{A}_{n, m}\nonumber \\ -R_{2}|\tilde{A}_{n, m}|^4\tilde{A}_{n, m}+R_{3}|\tilde{A}_{n, m}|^6\tilde{A}_{n, m}=0,
   \end{eqnarray}
   \begin{eqnarray}
  i\left(\frac{\partial}{\partial \tau}+\sigma \frac{\partial}{\partial \zeta}\right)\tilde{B}_{n, m}
  + \sigma \Delta \tilde{B}_{n, m} + \nonumber \\
   \qquad \qquad + \alpha_1 (\tilde{A}_{n, m}+\tilde{A}_{n+1, m}) +R_{1}|\tilde{B}_{n, m}|^2\tilde{B}_{n, m}\nonumber \\-R_{2}|\tilde{B}_{n, m}|^4\tilde{B}_{n, m}+R_{3}|\tilde{B}_{n, m}|^6\tilde{B}_{n, m}=0,
     \end{eqnarray}
     \begin{eqnarray}
   i\left(\frac{\partial}{\partial \tau}+\sigma \frac{\partial}{\partial \zeta}\right)\tilde{C}_{n, m}
  +\sigma \Delta\tilde{C}_{n, m} + \nonumber \\
  + \alpha_2 (\tilde{A}_{n, m}+\tilde{A}_{n, m+1}) +R_{1}|\tilde{C}_{n, m}|^2\tilde{C}_{n, m}\nonumber\\ -R_{2}|\tilde{C}_{n, m}|^4\tilde{C}_{n, m}+R_{3}|\tilde{C}_{n, m}|^6\tilde{C}_{n, m}=0.
  \end{eqnarray}
\end{subequations}
The next approximation is the zero mismatch ($\Delta=0$), which indicates that all the three waveguides of the unit cell posses the same propagation constant. Then the system
of equations takes the following forms,
\begin{subequations}
\label{eq:ABC:DEltazero:1}
\begin{eqnarray}
i\left(\frac{\partial}{\partial \tau}+ \frac{\partial}{\partial \zeta}\right)\tilde{A}_{n, m}
  + \alpha_1 (\tilde{B}_{n, m}+\tilde{B}_{n-1,m})\nonumber\\
 +\alpha_2 (\tilde{C}_{n, m}+\tilde{C}_{n, m-1})
   + R_{1}|\tilde{A}_{n, m}|^2\tilde{A}_{n, m}\nonumber\\-R_{2}|\tilde{A}_{n, m}|^4\tilde{A}_{n, m}+R_{3}|\tilde{A}_{n, m}|^6\tilde{A}_{n, m} =0,
   \end{eqnarray}
   \begin{eqnarray}
 i\left(\frac{\partial}{\partial \tau}+\sigma \frac{\partial}{\partial \zeta}\right)\tilde{B}_{n, m}
  \nonumber \\
 + \alpha_1 (\tilde{A}_{n, m}+\tilde{A}_{n+1, m}) +R_{1}|\tilde{B}_{n, m}|^2\tilde{B}_{n, m} \nonumber \\-R_{2}|\tilde{B}_{n, m}|^4\tilde{B}_{n, m}+R_{3}|\tilde{B}_{n, m}|^6\tilde{B}_{n, m}=0,
  \end{eqnarray}
  \begin{eqnarray}
  i\left(\frac{\partial}{\partial \tau}+\sigma \frac{\partial}{\partial \zeta}\right)\tilde{C}_{n, m}
    \nonumber \\
   + \alpha_2 (\tilde{A}_{n, m}+\tilde{A}_{n,
  m+1}) +R_{1}|\tilde{C}_{n, m}|^2\tilde{C}_{n, m}\nonumber \\-R_{2}|\tilde{C}_{n, m}|^4\tilde{C}_{n, m}+R_{3}|\tilde{C}_{n, m}|^6\tilde{C}_{n, m}=0.
\end{eqnarray}
\end{subequations}
\begin{figure*}[t]
 	\includegraphics[width=.33\linewidth]{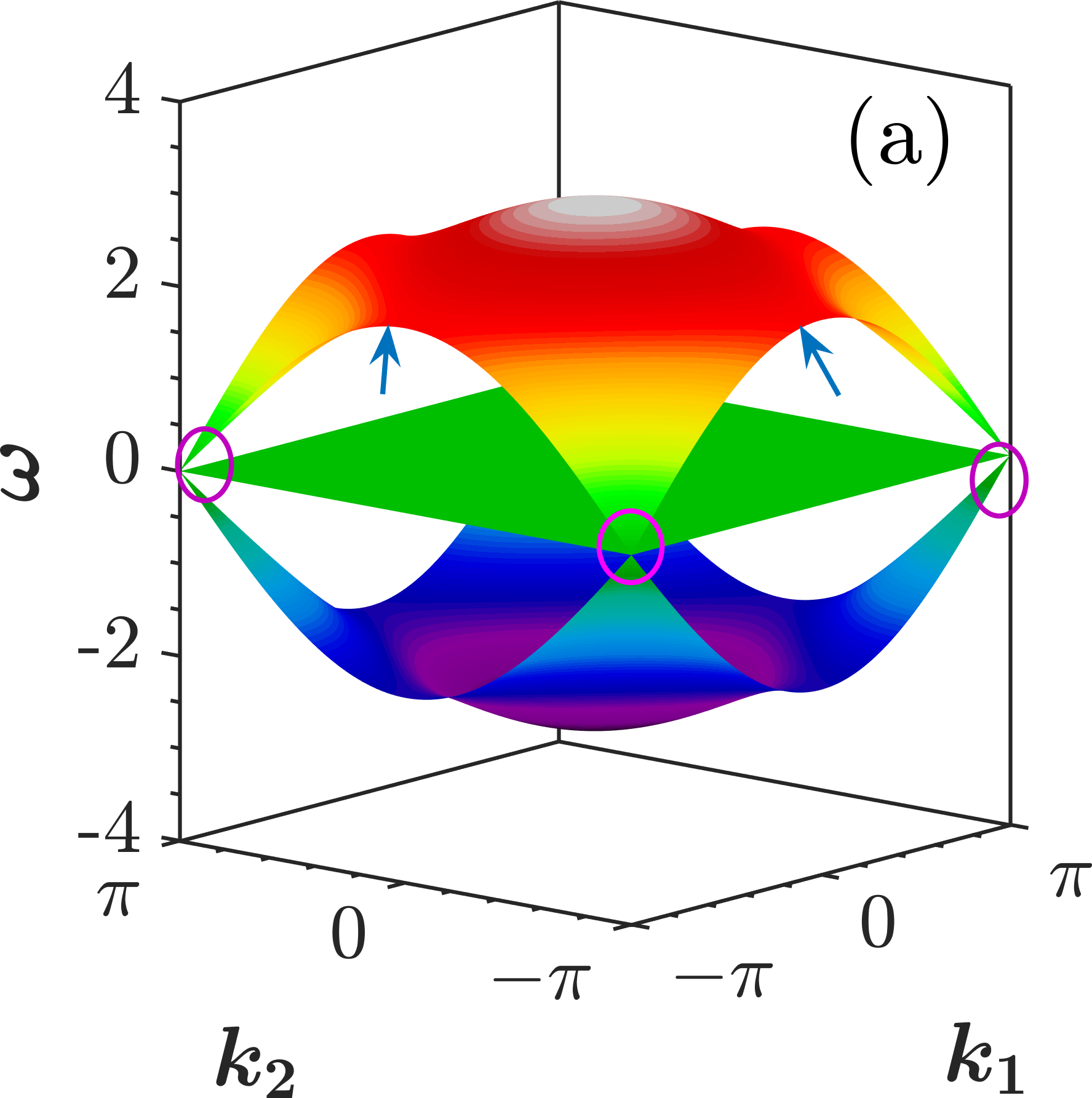}
 	\includegraphics[width=.33\linewidth]{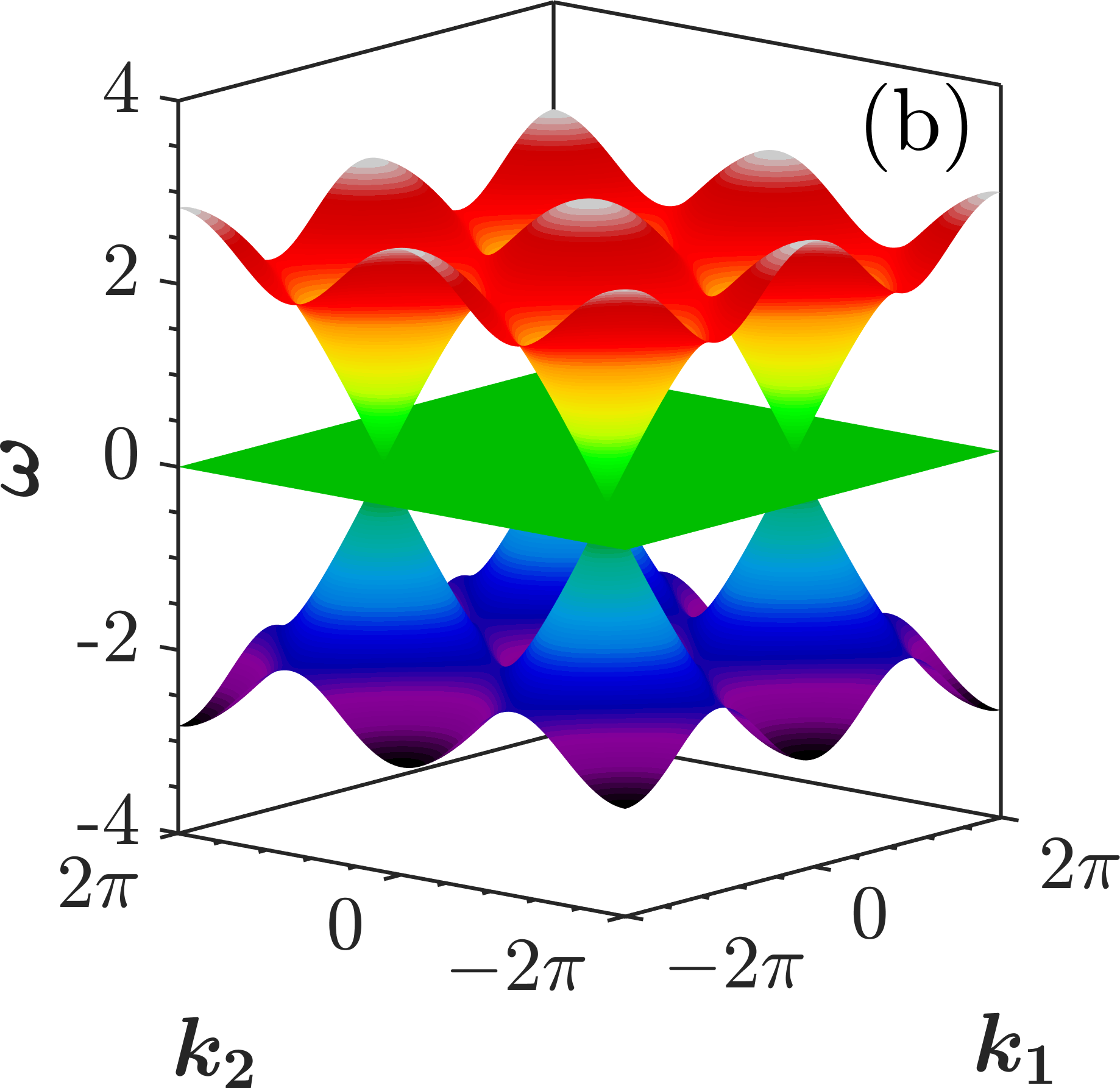}\includegraphics[width=.28\linewidth]{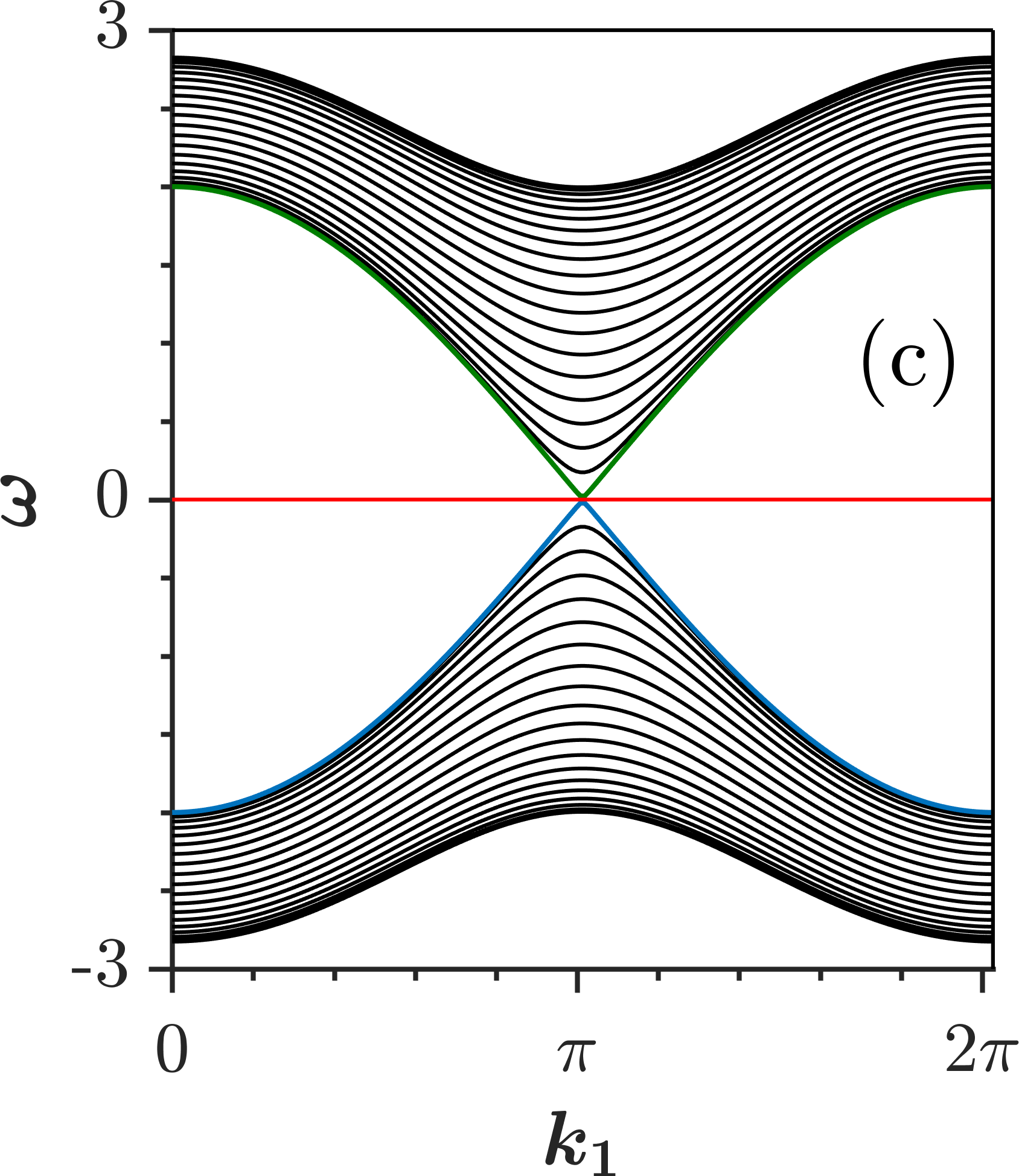}
 	\caption{(Color online.) The linear dispersion relation of Lieb lattice featuring two-dimensional  waveguide arrays. (a) shows the three bands in the first Brillouin zone ($k_{1,2}\in [-\pi, \pi]$) and (b) depicts the same in the second Brillouin regime ($k_{1,2} \in [-2\pi, 2\pi]$). Here the flat band (drawn in green color) is separated by two dispersive (upper and lower) bands, indicated by the color combinations of yellow, red and blue, purple respectively. (c) portrays a band structure of $n=20$ unit cells with $k_{1,2} \in [0, 2\pi]$. The system parameters are $\alpha_1=\alpha_2=l=1$, and $R_1=R_2=R_3\equiv R=0$.}
 	\label{lindis}
 \end{figure*}
Let us now discuss the dispersion relations of the waves which are governed by Eqs.  (\ref{eq:ABC:DEltazero:1}).
Consider the quasi-harmonic waves
\begin{subequations}
\begin{eqnarray}
   \tilde{A}_{mn}=A_0 e^{-i\omega \tau +ik_z \zeta+ ik_1n+ik_2m},
   \end{eqnarray}
   \begin{eqnarray}
 \tilde{B}_{mn}=B_0 e^{-i\omega \tau +ik_z \zeta+ ik_1n+ik_2m},
     \end{eqnarray}
     \begin{eqnarray}
 \tilde{C}_{mn}=C_0 e^{-i\omega \tau +ik_z \zeta+ ik_1n+ik_2m},
  \end{eqnarray}
\end{subequations}
where $k_z$ is a small correction to the constant of propagation along the waveguide, $k_1 = k_x l$
and $k_2=k_y l$ are normalized wave-numbers and $l$ is the lattice parameter. The quantities $k_x$ and $k_y$ are the quasi (Bloch)
momenta of the 2D Lieb lattice. Substitution of these expressions in Eq. (\ref{eq:ABC:DEltazero:1}) results in the following
system of algebraic equations,
\begin{subequations}
\label{sysqw}
\begin{eqnarray}
  (\omega-k_z+f_1)A_0+\kappa_1^*B_0+\kappa_2^*C_0=0,
   \end{eqnarray}
   \begin{eqnarray}
  \kappa_1A_0+(\omega-\sigma k_z+f_2)B_0=0,
     \end{eqnarray}
     \begin{eqnarray}
  \kappa_2A_0+(\omega-\sigma k_z+f_3)C_0=0,
  \end{eqnarray}
\end{subequations}
where the nonlinearity contributions in the dispersion relations are
\begin{subequations}
\begin{eqnarray}
  f_1=R_{1}|A_0|^2-R_{2}|A_0|^4+R_{3}|A_0|^6,
   \end{eqnarray}
   \begin{eqnarray}
   f_2=R_{1}|B_0|^2-R_{2}|B_0|^4+R_{3}|B_0|^6,
     \end{eqnarray}
     \begin{eqnarray}
  f_3=R_{1}|C_0|^2-R_{2}|C_0|^4+R_{3}|C_0|^6.
  \end{eqnarray}
\end{subequations}
In Eq. (\ref{sysqw}) we have also introduced of the normalized wave-numbers,
\begin{subequations}
\begin{eqnarray}
  \kappa_1=\alpha_1(1+e^{ik_1}),
   \end{eqnarray}
   \begin{eqnarray}
   \kappa_2=\alpha_2(1+e^{ik_2}).
     \end{eqnarray}
   \end{subequations}

Non-zero solutions of Eq. (\ref{sysqw}) exist if the determinant associated with this system of equations is equal to
zero. It results in the dispersion equation,
\begin{eqnarray}
\label{disp1}
(\omega-k_z+f_1)(\omega-\sigma k_z+f_2)(\omega-\sigma k_z+f_3)\nonumber \\-(\omega-\sigma k_z+f_2)|\kappa_2|^2-(\omega-\sigma k_z+f_3)|\kappa_1|^2=0.
\end{eqnarray}
The general solution to this equation is given in Appendix 1. For a particular case $f_2 = f_3$,
 the dispersion relation $\omega=\omega(\kappa_1, \kappa_2, k_z; A_0,B_0,C_0)$  becomes a factorized one, which results in the following relations,
\begin{subequations}
\label{dis}
\begin{eqnarray}
\label{dis1}
 \omega=\sigma k_z-f_2,
   \end{eqnarray}
   \begin{eqnarray}
 (\omega- k_z+f_1)(\omega-\sigma k_z+f_2)-(|\kappa_1|^2+|\kappa_2|^2)=0.
 \end{eqnarray}
 \label{Eqdis}
 \end{subequations}
 The first branch of the dispersion relation, i.e. Eq. (\ref{dis1}), corresponds to diffractionless wave propagation (as $d\omega/dk$ is independent of the wavenumber). At $\sigma = 1$ it is a forward wave, and at $\sigma=-1$ it is a backward one. Note that the backward propagation of the wave is due to the negative refraction in the metamaterial.
 \begin{figure*}[t]
 	\includegraphics[width=.28\linewidth]{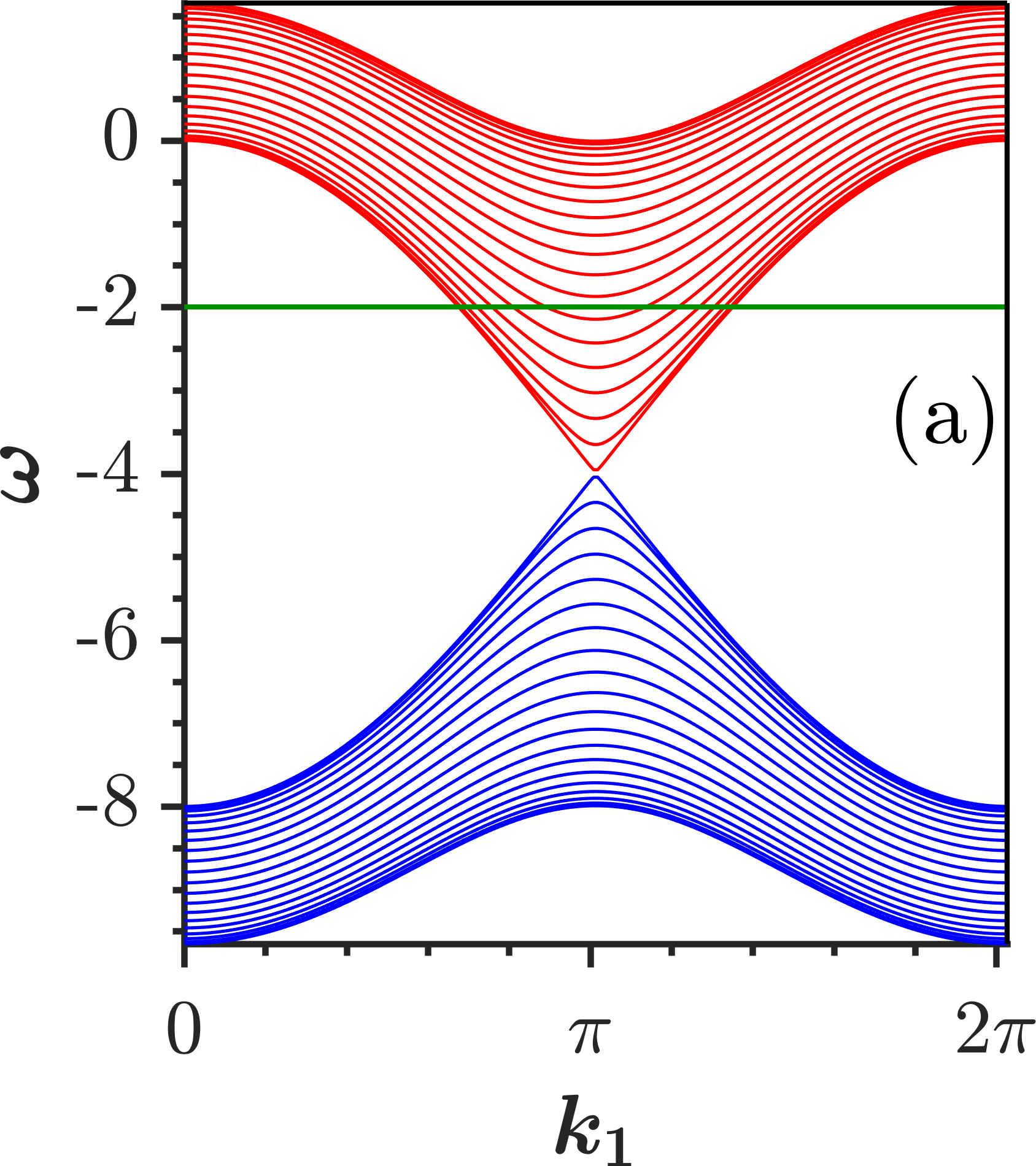}
 	\includegraphics[width=.28\linewidth]{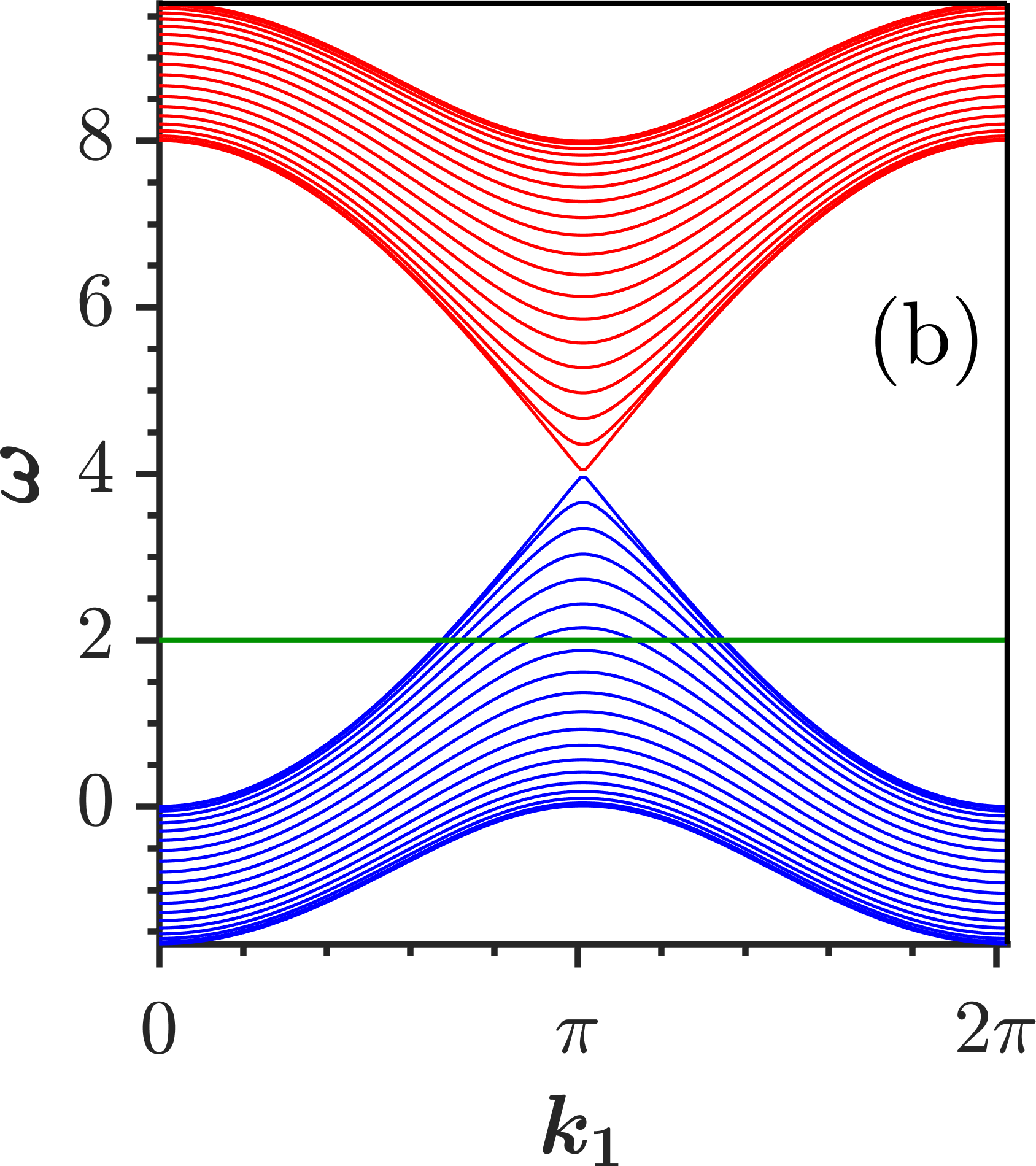}\includegraphics[width=.28\linewidth]{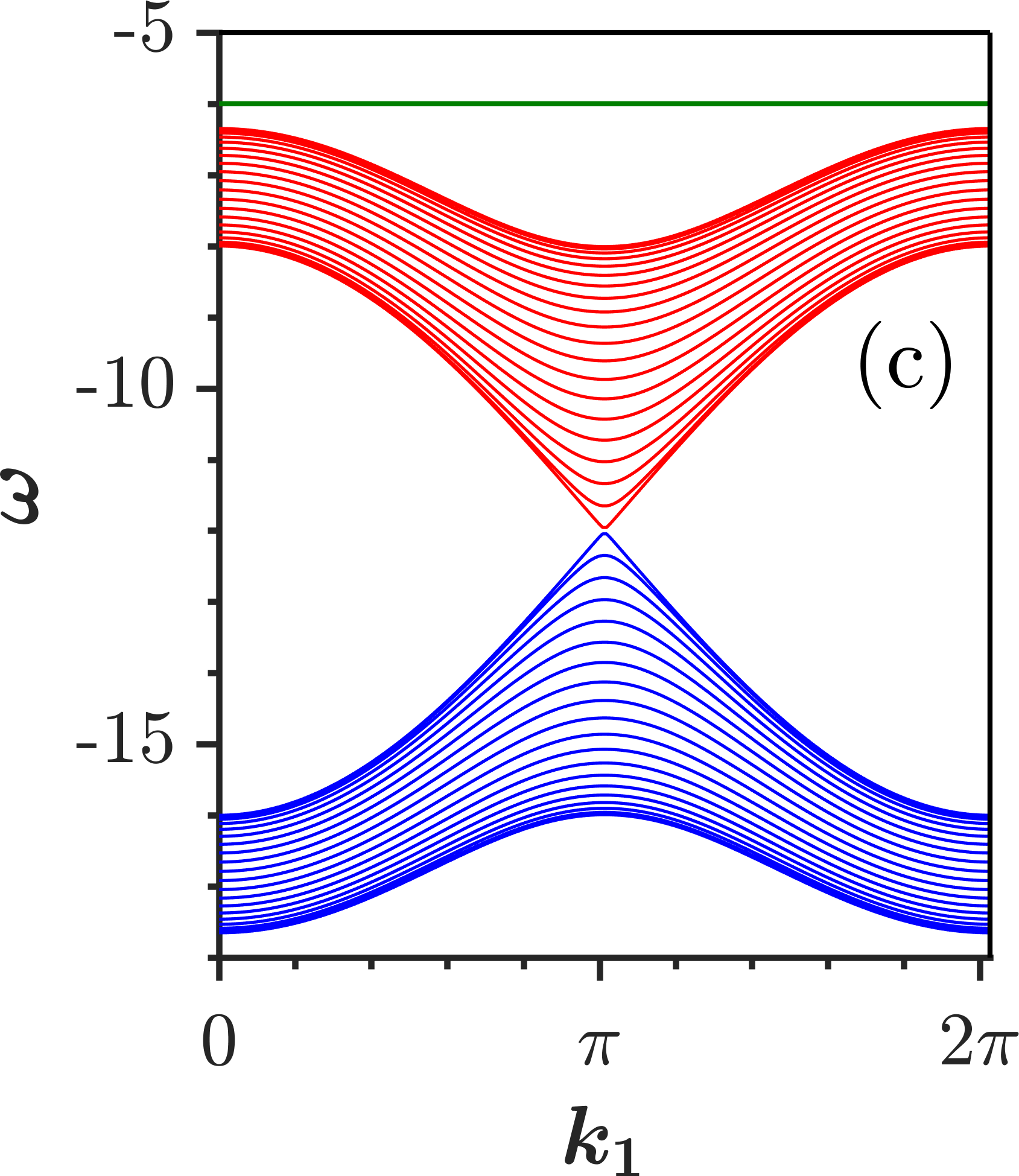}\\
 	\includegraphics[width=.3\linewidth]{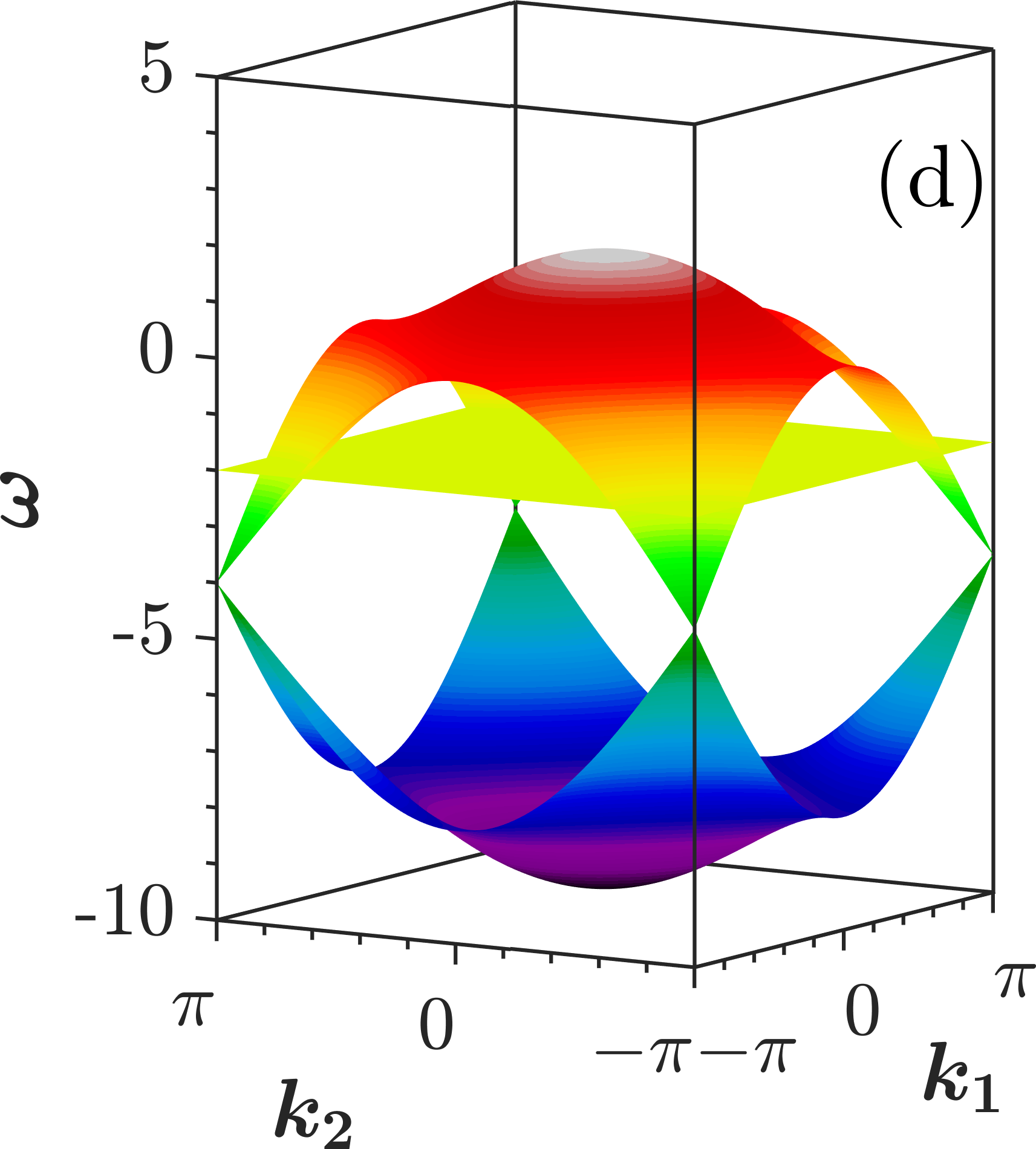}
 	\includegraphics[width=.3\linewidth]{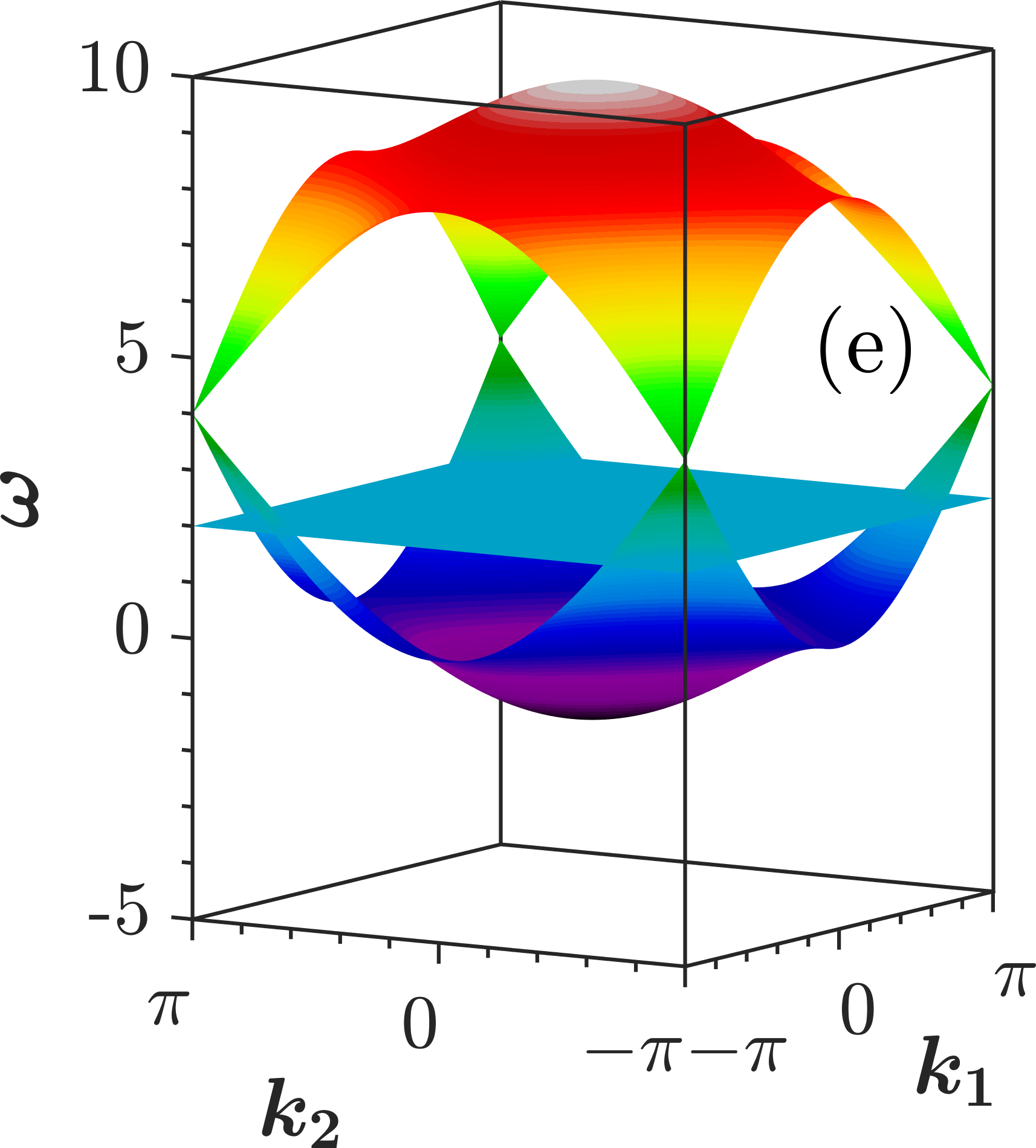}\includegraphics[width=.3\linewidth]{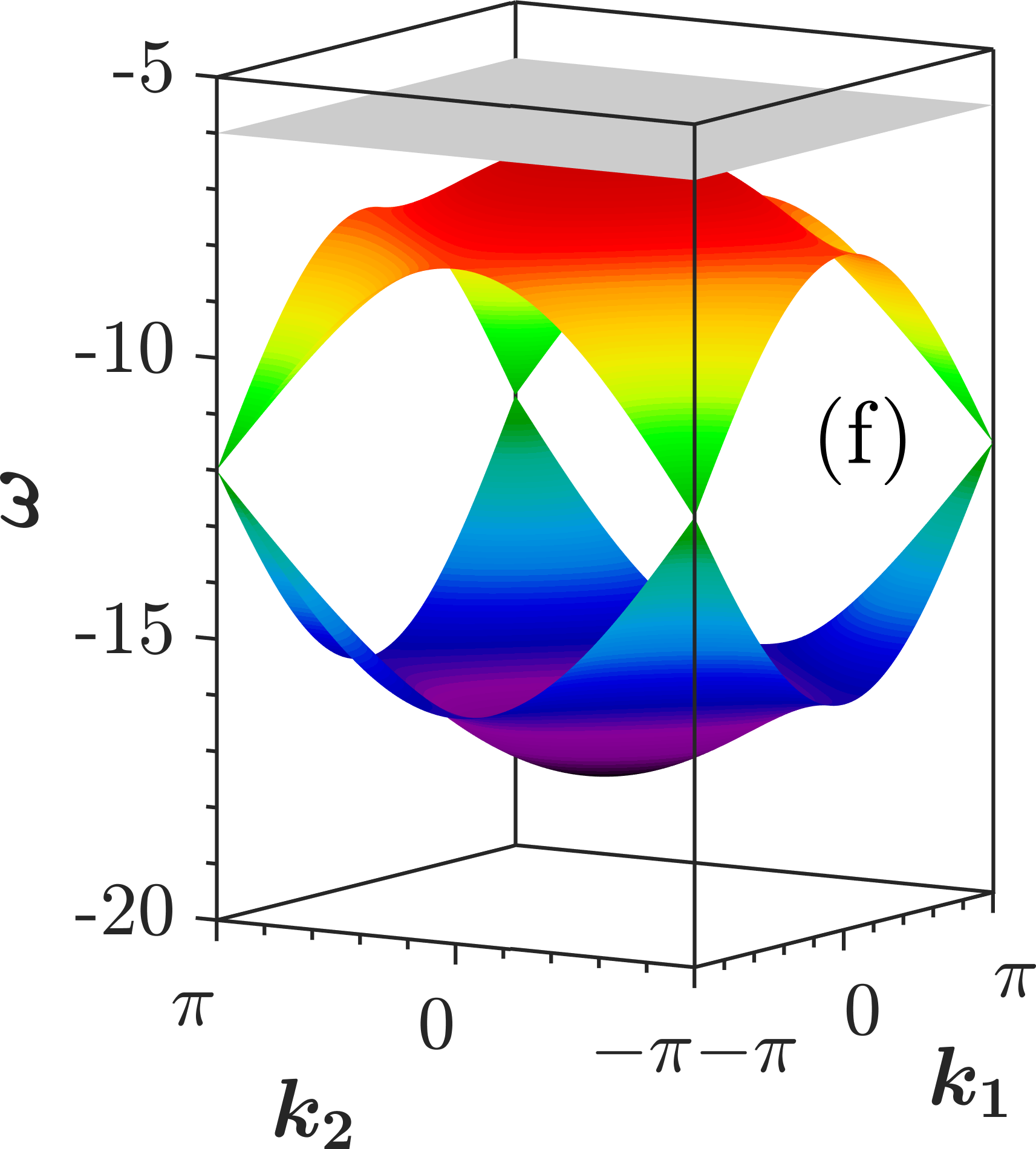}
 	\caption{(Color online.) The nonlinear dispersion relation of Lieb lattice with 20 unit cells ($n=20$) for the same parameters as in Fig. \ref{lindis} except, (a) $R_2=R_3=0, R_1=1$, (b) $R_1=R_2=1, R_3=0$, and (c) $R_1=R_2=1=R_3=1$ with $|A_0|^2=|B_0|^2=|C_0|^2=1$. Bottom panels indicate the corresponding three dimensional structure for a finite Lieb lattice.}
 	\label{nldis}
 \end{figure*}

 We first present the dispersion characteristics of Eq. (\ref{Eqdis}) for the linear system.
 Then for a linear system $R_1=R_2=R_3\equiv R=0$ and the case is depicted in Fig. \ref{lindis}. It is to be noted that to sketch the dispersion relations, we have considered the frequency of the Lieb lattice as a function of the two dimensional Bloch wave vectors $k_1$ and $k_2$ by neglecting the small correction in the propagation constant ($k_z=0$).  One can clearly notice from Fig. \ref{lindis} drawn in the  first Brillouin zone ($k_{1,2} \in [-\pi, \pi]$) that the system supports the typical dispersion curves with three energy bands including a perfectly flat band, which is identical to the energy bands observed in photonic Lieb lattices \cite{Mukherjee:Spracklen:15a, ppbelicev} and Kagome lattices in addition to the  two dispersive bands \cite{liang}. Such a characteristics where all the three bands get into contact with the symmetric band, that is the flat band is known as the particle-hole symmetry analogous to the quantum version. Also, this  flat band is a manifestation of degenerate state meaning that it is static and will not contribute to any transport of localized state. This implies that the localized states of flat bands are diffractionless since their group-velocity is zero.  When the dispersion relation is plotted in the second Brillouin zone ($k_{1,2} \in [-2\pi, 2\pi]$), quite a number of unique features of bands gets revealed. For instance, one can observe Dirac cones (marked with circles in Fig. \ref{lindis}(a)) of the conical dispersive bands intersect with the flat band at the corner of the first Brillouin zone. Also, Tamm-like edge states of dispersive bands result in a van Hove singularity in the given three bands (see blue colored arrow marks in Fig. \ref{lindis}(a)). Further, the band structure shown in Fig. \ref{lindis}(c) (obtained for 20 unit cells ($n=20$)) clearly reveals the topologically protected solid edge states (drawn in green color curve) in addition to the typical bulk states (black lined curves).

 The nonlinear dispersion relation is shown in Fig. \ref{nldis}. As the cubic nonlinearity acts alone,  the flat band shifts towards low (negative) frequency from zero one as depicted in Figs. \ref{nldis}(a) and \ref{nldis}(d) while both the dispersion (conical) bands get shifted towards the negative frequency from the positive side. On the other hand, if the quintic nonlinearity is invoked along with the cubic one, one can observe that the shift of flat band is quite opposite to the cubic one. When the system includes the septimal nonlinearity, besides the cubic and quintic ones, the flat band and dispersive bands completely get drifted towards the negative frequency. Hence it is clear that when we include higher order nonlinearities such as cubic, quintic and septimal ones, the degeneracy of the flat band is even reduced and located on the top of conical bands. Thus, one can conclude that the inclusion of nonlinearity shifts the flat band from zero frequency to higher as well lower values depending upon the type of nonlinearities. In other words, the high intensity optical light alters the location of degenerate and dispersive bands. These ramifications clearly suggest the possibility of optically controlling the band structure of a Lieb waveguide array.
 \par
In the linear regime  with the following approximations,  Eqs. (\ref{eq:ABC:DEltazero:1})
have particular solutions, which are written as
\begin{eqnarray}
  &&(\mathrm{A}) ~~  \tilde{A}_{n, m}=0,\quad \alpha_1  \tilde{B}_{n, m} =-\alpha_2  \tilde{C}_{n,m-1}, \nonumber \\
  && \qquad \qquad  \alpha_1  \tilde{B}_{n-1, m} =-\alpha_2  \tilde{C}_{n, m},  \label{eq:2D:sqlat:nl:A1} \\
  && (\mathrm{B}) ~~~  \tilde{A}_{n, m}=0,\quad \alpha_1  \tilde{B}_{n, m} =-\alpha_1  \tilde{B}_{n-1, m}, \nonumber\\
  && \qquad \qquad \alpha_2  \tilde{C}_{n, m} =-\alpha_2  \tilde{C}_{n, m-1}, \label{eq:2D:sqlat:nl:A2} \\
  && (\mathrm{C})  \tilde{A}_{n, m}=(-1)^{n+m} \tilde{A}, \quad  \tilde{B}_{n,m}=(-1)^{n+m} \tilde{B}, \nonumber\\
  && \qquad \qquad   \tilde{C}_{n, m}=(-1)^{n+m} \tilde{C}.  \label{eq:2D:sqlat:nl:A3}
\end{eqnarray}
It is worthwhile to mention that the diffractionless propagation of electromagnetic waves in the two dimensional
waveguide arrays under consideration is described by these solutions
in such a linear approximation. Similar behavior in the nonlinear case will be considered in the next section.
\section{The diffractionless solutions and their stability}
\begin{figure*}[t]
	\subfigure[$R_1=1$ and  $R_2=R_3=0$]{\label{MI1A}\includegraphics[width=0.32\linewidth]{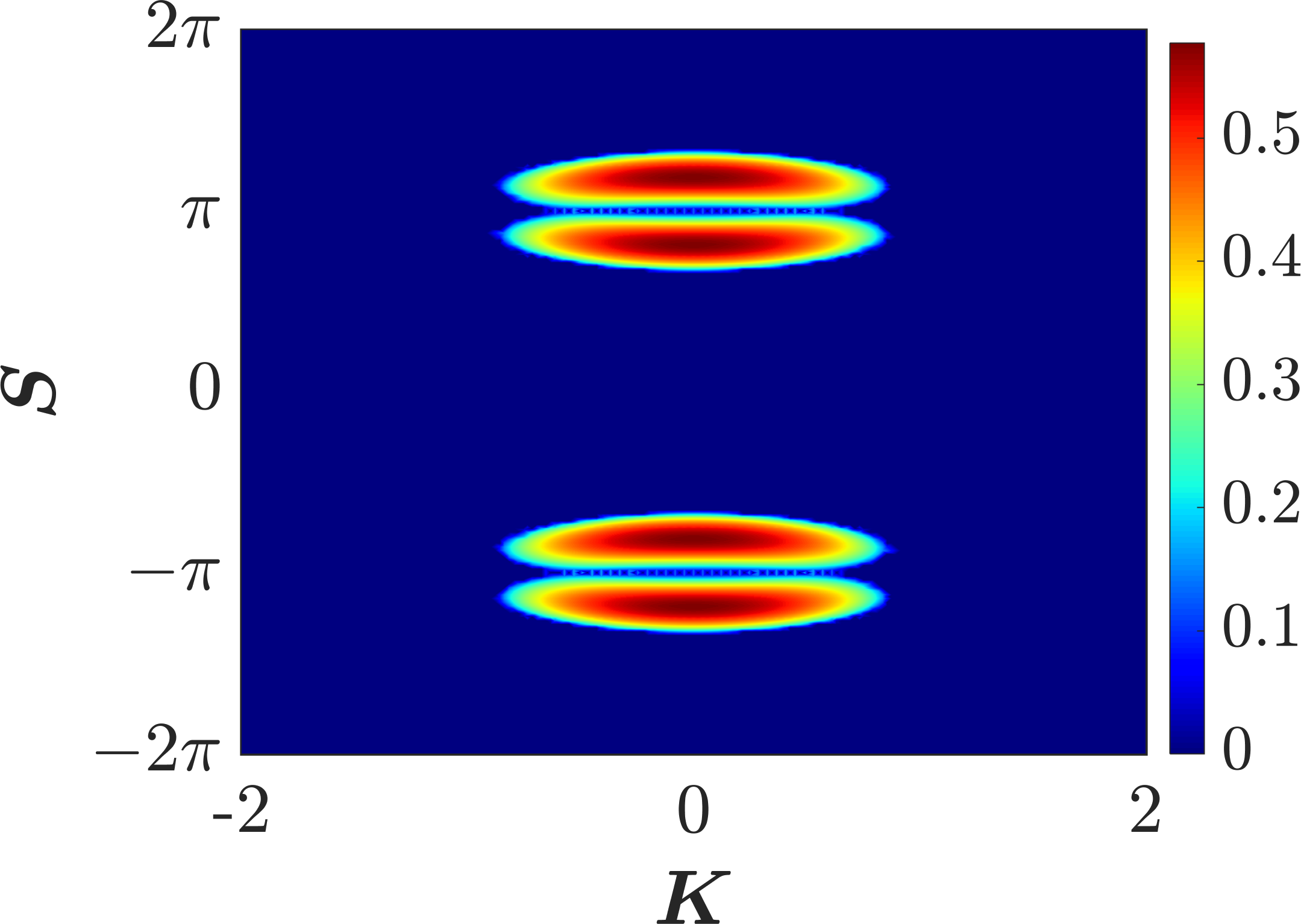}}
	\subfigure[$R_1=R_2=R_3=1$]{\label{MI1B}\includegraphics[width=0.32\linewidth]{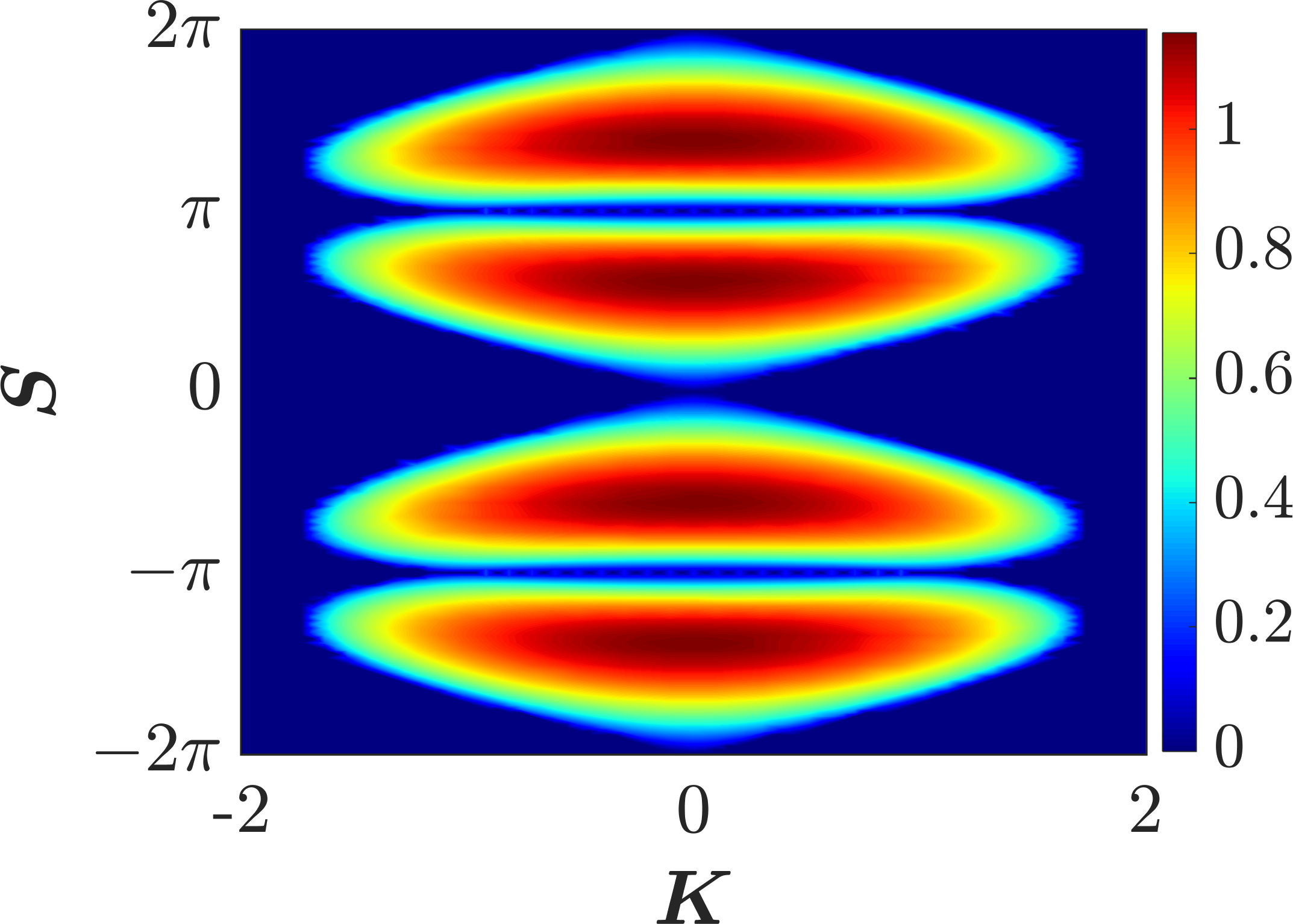}}
	\subfigure[$R_1=-R_2=1$ and $R_3=0$]{\label{MI1C}\includegraphics[width=0.32\linewidth]{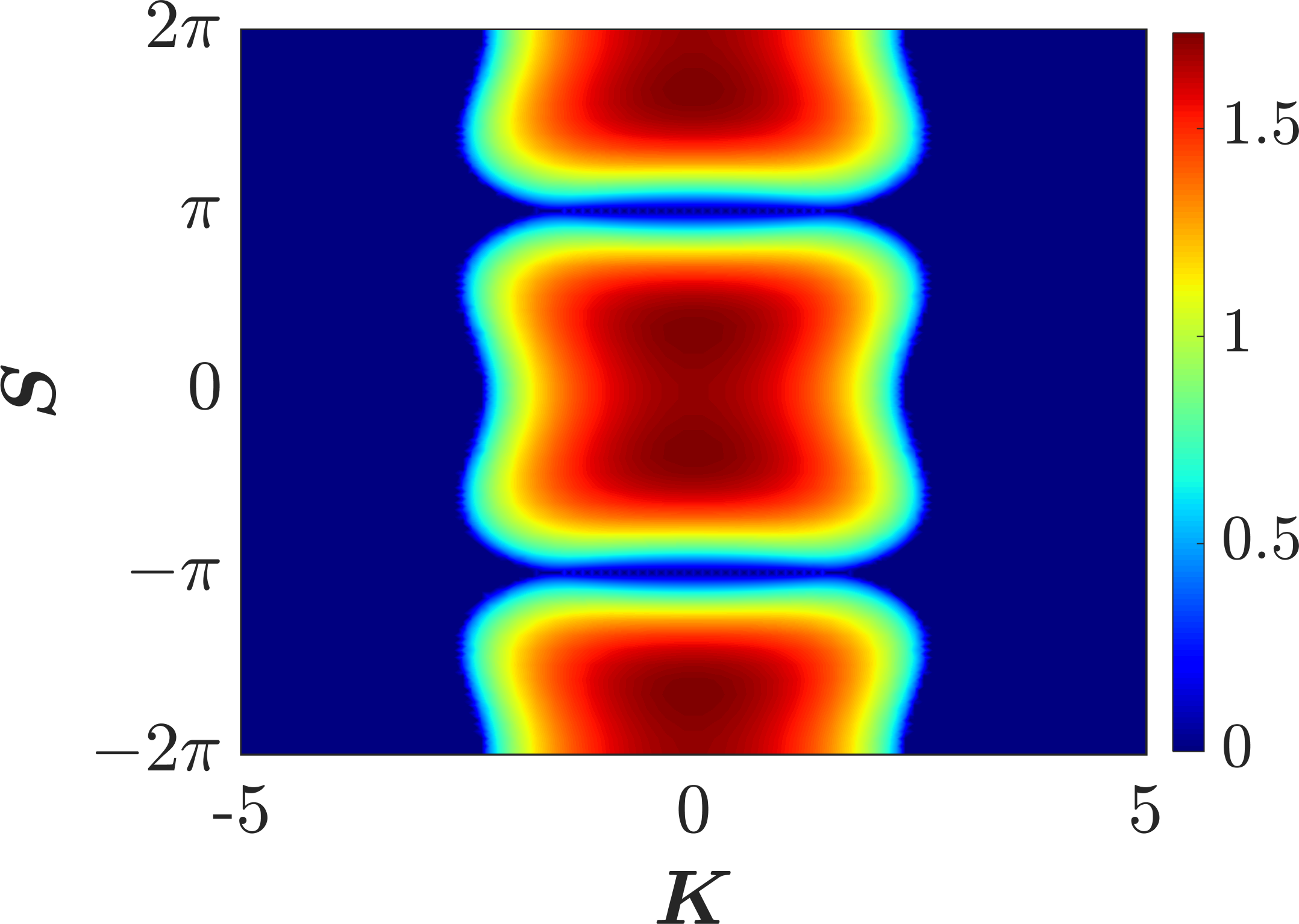}}
	\subfigure[$R_1=-R_2=R_3=1$]{\label{MI1D}\includegraphics[width=0.32\linewidth]{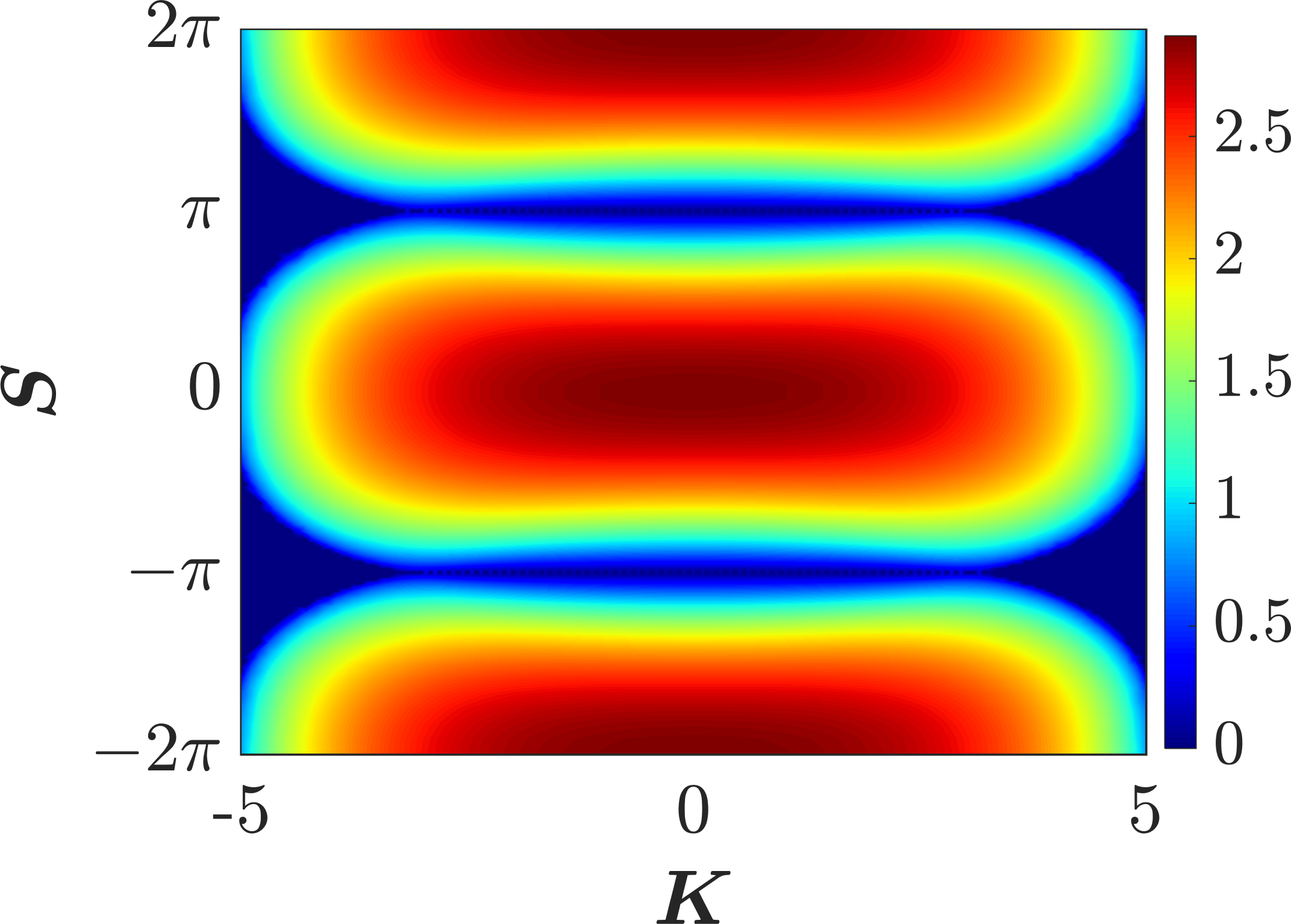}}
	\subfigure[$R_1=R_2=-R_3=1$]{\label{MI1E}\includegraphics[width=0.32\linewidth]{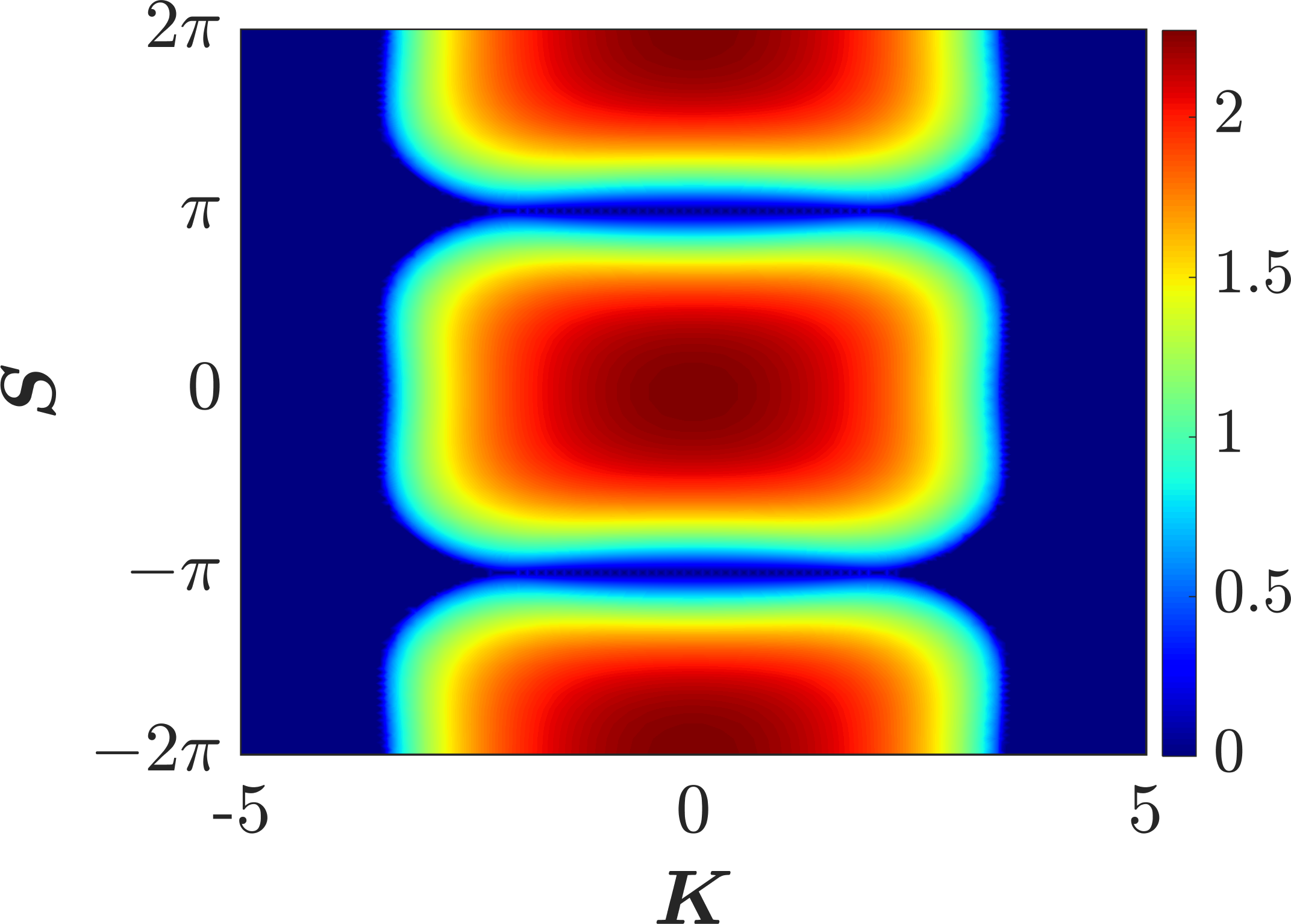}}
	\caption{(Color online.) The MI gain spectra versus $K$ as a function of  $p=q=S$ for different combinations of nonlinear coefficients.}
	\label{MI1}
\end{figure*}
\noindent To find the nonlinear analogy of Eqs. (\ref{eq:2D:sqlat:nl:A1})--(\ref{eq:2D:sqlat:nl:A3}), let us suppose that $\tilde{A}_{n, m}$ is zero at all $\zeta$ and $\tau$. Then the system of Eqs. (\ref{eq:ABC:DEltazero:1}) is reduced to following equations,
\begin{subequations}
\begin{eqnarray}
   i\left(\frac{\partial}{\partial \tau}+\sigma \frac{\partial}{\partial \zeta}\right)\tilde{B}_{n, m}
    + R_{1}|\tilde{B}_{n, m}|^2\tilde{B}_{n, m}\nonumber\\-R_{2}|\tilde{B}_{n, m}|^4\tilde{B}_{n, m}+R_{3}|\tilde{B}_{n, m}|^6\tilde{B}_{n, m}=0,  \label{eq:ABC:DEltazero:Bn1}
 \end{eqnarray}
 \begin{eqnarray}
 i\left(\frac{\partial}{\partial \tau}+\sigma \frac{\partial}{\partial \zeta}\right)\tilde{C}_{n, m}
    + R_{1}|\tilde{C}_{n, m}|^2\tilde{C}_{n, m}\nonumber\\-R_{2}|\tilde{C}_{n, m}|^4\tilde{C}_{n, m}+R_{3}|\tilde{C}_{n, m}|^6\tilde{C}_{n, m}=0.\label{eq:ABC:DEltazero:Cn1}
\end{eqnarray}
\end{subequations}
The homogeneous solution of these equations have the following form
\begin{eqnarray}\label{eq:ABC:DEltazero:BnCn:1}
    \tilde{A}_{n, m}=0,\quad \tilde{B}_{n, m}=B_0e^{i(R_{1}|B_0|^2-R_{2}|B_0|^4+R_{3}|B_0|^6)\zeta}, \nonumber  \\ \tilde{C}_{n, m}=C_0e^{i(R_{1}|B_0|^2-R_{2}|B_0|^4+R_{3}|B_0|^6)\zeta},
\end{eqnarray}
where the constraints $\alpha_2=\alpha_1=1$ and $|C_0|=|B_0|$ are used.

 To study the stability of the associated solutions
(\ref{eq:ABC:DEltazero:BnCn:1}),  the appropriately perturbed amplitudes must be
introduced. For instance we choose,
\begin{eqnarray}
   && \tilde{A}_{n, m} =a_{n, m}e^{i\varphi}, \quad \tilde{B}_{n,m} =
(B_0+b_{n,m})e^{i\varphi},\nonumber  \\
   && \tilde{C}_{n,m}
=(-B_0+c_{n,m})e^{i\varphi},\nonumber
\end{eqnarray}
where $\partial\varphi/\partial\xi = R_1B_0^2-R_2B_0^4+R_3B_0^6$  .
The linearization  of the system of Eqs. (\ref{eq:ABC:DEltazero:1}) results in the following equations
\begin{subequations}
\begin{eqnarray}
 i\frac{\partial a_{n m}}{\partial \eta} +a_{n m} (R_1 B_0^2 -R_2 B_0^4 + R_3 B_0^6 )\nonumber \\
 + (b_{n, m}+b_{n-1, m}) +(c_{n,m}+c_{n, m-1}) =0,
 \end{eqnarray}
 \begin{eqnarray}
  i\frac{\partial b_{n m}}{\partial \xi} +(a_{n m}+a_{n+1, m})+ R_1 B_0^2 \left(
b_{nm}^{*}+ b_{nm} \right)\nonumber \\-R_2 B_0^4 \left(
b_{nm}^{*}+ b_{nm} \right)+R_3 B_0^6 \left(
b_{nm}^{*}+ b_{nm} \right)=0,
\end{eqnarray}
\begin{eqnarray}
 i\frac{\partial c_{n m}}{\partial \xi} +(a_{n m}+a_{n, m+1})+ R_1 B_0^2 \left(
c_{nm}^{*}+ c_{nm} \right)\nonumber \\-R_2 B_0^4 \left(
c_{nm}^{*}+ c_{nm} \right)+R_3 B_0^6 \left(
c_{nm}^{*}+ c_{nm} \right)=0,
\end{eqnarray}
 \label{eq:2D:sqlat:nl:FB:2}
\end{subequations}
where
$$
\frac{\partial }{\partial \eta} = \left(\frac{\partial}{\partial
\tau}+ \frac{\partial}{\partial \zeta}\right), \quad \frac{\partial
}{\partial \xi} = \left(\frac{\partial}{\partial \tau}+\sigma
\frac{\partial}{\partial \zeta}\right).
$$

Disinflation of the system of linear equations can be done by using
the Fourier transformations,
\begin{eqnarray}
  && a_{nm} = \sum_{p,q}\left(a_{pq}e^{ipn+iqm}+ \tilde{a}_{pq}e^{-ipn-iqm}\right), \nonumber \\
  && b_{nm} = \sum_{p,q}\left(b_{pq}e^{ipn+iqm}+ \tilde{b}_{pq}e^{-ipn-iqm}\right), \nonumber \\
  && c_{nm} = \sum_{p,q}\left(c_{pq}e^{ipn+iqm}+ \tilde{c}_{pq}e^{-ipn-iqm}\right).\nonumber
\end{eqnarray}
Equations for the Fourier amplitudes $a_{pq}$, $b_{pq}$, $c_{pq}$,
$\bar{a}_{pq}$, $\bar{b}_{pq}$ and $\bar{c}_{pq}$ follow from
Eqs. (\ref{eq:2D:sqlat:nl:FB:2}) as,
\begin{subequations}
\label{eq:2D:sqlat:nl:FB:3}
\begin{eqnarray}
  i\frac{\partial a_{pq}}{\partial \eta} +a_{pq}(R_1 B_0^2-R_2 B_0^4+R_3 B_0^6)
 + \kappa_1 b_{pq} + \kappa_2c_{pq} =0,
 \end{eqnarray}
 \begin{eqnarray}
  i\frac{\partial b_{pq}}{\partial \xi} + \kappa_1^{*}a_{pq}+ R_1 B_0^2 \left(
b_{pq}+ \bar{b}_{pq}^{*} \right)\nonumber \\-R_2 B_0^4 \left(
b_{pq}+ \bar{b}_{pq}^{*} \right)+R_3 B_0^6 \left(
b_{pq}+ \bar{b}_{pq}^{*} \right)=0,
\end{eqnarray}
\begin{eqnarray}
i\frac{\partial c_{pq}}{\partial \xi} + \kappa_2^{*}a_{pq}+ R_1 B_0^2 \left(
c_{pq}+ \bar{c}_{pq}^{*} \right)\nonumber \\-R_2 B_0^4 \left(
c_{pq}+ \bar{c}_{pq}^{*} \right)+R_3 B_0^6 \left(
c_{pq}+ \bar{c}_{pq}^{*} \right)=0,
\end{eqnarray}
\begin{eqnarray}
 i\frac{\partial \bar{a}_{pq}}{\partial \eta} +\bar{a}_{pq}(R_1 B_0^2-R_2 B_0^4+R_3 B_0^6 )
 + \kappa_1^{*} \bar{b}_{pq} + \kappa_2^{*}\bar{c}_{pq} =0,
 \end{eqnarray}
 \begin{eqnarray}
  i\frac{\partial \bar{b}_{pq}}{\partial \xi} + \kappa_1 \bar{a}_{pq}+ R_1 B_0^2 \left(
b_{pq}^{*} + \bar{b}_{pq}\right)\nonumber \\-R_2 B_0^4 \left(
b_{pq}^{*} + \bar{b}_{pq}\right)+R_3 B_0^6 \left(
b_{pq}^{*} + \bar{b}_{pq}\right)=0,
\end{eqnarray}
 \begin{eqnarray}
  i\frac{\partial \bar{c}_{pq}}{\partial \xi} + \kappa_2 \bar{a}_{pq}+ R_1 B_0^2 \left(
c_{pq}^{*}+ \bar{c}_{pq} \right)\nonumber \\- R_2 B_0^4 \left(
c_{pq}^{*}+ \bar{c}_{pq} \right)+ R_3 B_0^6 \left(
c_{pq}^{*}+ \bar{c}_{pq} \right)=0.
\end{eqnarray}
\end{subequations}
Here, the coefficient functions are denoted as
\begin{eqnarray}
   && \qquad \kappa_1=\kappa(p),\quad \kappa_2=\kappa(q), \nonumber  \\
   && \kappa(k) = 1+e^{-ik} = 2\cos(k/2)e^{-ik/2}. \nonumber
\end{eqnarray}

In order to solve the above system of six linear differential
equations, we assume the following plane wave ansatz,
\begin{eqnarray}
&& a_{pq}= a e^{iK\zeta -i\Omega \tau}, \quad  \bar{a}_{pq}= \bar{a} e^{-iK\zeta + i\Omega \tau}, \nonumber \\
&& b_{pq}= b e^{iK\zeta -i\Omega \tau}, \quad \bar{b}_{pq}= \bar{b} e^{-iK\zeta + i\Omega \tau},\nonumber \\
&& c_{pq}= c e^{iK\zeta -i\Omega \tau }, \quad \bar{c}_{pq}= \bar{c}
e^{-iK\zeta + i\Omega \tau}.\nonumber
\end{eqnarray}

Substituting these expressions in Eq. (\ref{eq:2D:sqlat:nl:FB:3}), we
obtain a set of linearly coupled algebraic equations for $a$, $\bar{a}$, b, $\bar{b}$, c and $\bar{c}$. This
set has nontrivial solutions only when the 6x6 determinant
formed by the coefficient matrix vanishes as given below:
$$ \left(
   \begin{array}{cccccc}
     \epsilon_{11}&\epsilon_{12}&\epsilon_{13}&\epsilon_{14}&\epsilon_{15}&\epsilon_{16}\\
     \epsilon_{21}&\epsilon_{22}&\epsilon_{23}&\epsilon_{24}&\epsilon_{25}&\epsilon_{26}\\
     \epsilon_{31}&\epsilon_{32}&\epsilon_{33}&\epsilon_{34}&\epsilon_{35}&\epsilon_{36}\\
     \epsilon_{41}&\epsilon_{42}&\epsilon_{43}&\epsilon_{44}&\epsilon_{45}&\epsilon_{46}\\
     \epsilon_{51}&\epsilon_{52}&\epsilon_{53}&\epsilon_{54}&\epsilon_{55}&\epsilon_{56}\\
     \epsilon_{61}&\epsilon_{62}&\epsilon_{63}&\epsilon_{64}&\epsilon_{65}&\epsilon_{66}\\

   \end{array}
 \right)\left(
          \begin{array}{c}
            a\\
            b\\
            c\\
            \bar{a}\\
            \bar{b}\\
            \bar{c}\\

          \end{array}
        \right)=0,
 $$

where $\epsilon_{11}=-K+\Omega+R_1 B_0^2-R_2 B_0^4+R_3 B_0^6$, $\epsilon_{12}=\kappa_1$, $\epsilon_{13}=\kappa_2$, $\epsilon_{14}=0$, $\epsilon_{15}=0$
$\epsilon_{16}=0$, $\epsilon_{21}=\kappa_1^{*}$, $\epsilon_{22}=-\sigma K +\Omega+R_1 B_0^2-R_2 B_0^4+R_3 B_0^6$, $\epsilon_{23}=0$, $\epsilon_{24}=0$,
$\epsilon_{25}=R_1 B_0^2-R_2 B_0^4+R_3 B_0^6$, $\epsilon_{26}=0$, $\epsilon_{31}=\kappa_2^{*}$, $\epsilon_{32}=0$, $\epsilon_{33}=-\sigma K +\Omega+R_1 B_0^2-R_2 B_0^4+R_3 B_0^6$,
$\epsilon_{34}=0$, $\epsilon_{35}=0$, $\epsilon_{36}=R_1 B_0^2-R_2 B_0^4+R_3 B_0^6$, $\epsilon_{41}=0$, $\epsilon_{42}=0$, $\epsilon_{43}=0$,
$\epsilon_{44}=K-\Omega+R_1 B_0^2-R_2 B_0^4+R_3 B_0^6$, $\epsilon_{45}=\kappa_1^{*}$, $\epsilon_{46}=\kappa_2^{*}$, $\epsilon_{51}=0$, $\epsilon_{52}=R_1 B_0^2-R_2 B_0^4+R_3 B_0^6$,
$\epsilon_{53}=0$, $\epsilon_{54}=\kappa_1$, $\epsilon_{55}=\sigma K -\Omega+R_1 B_0^2-R_2 B_0^4+R_3 B_0^6$, $\epsilon_{56}=0$, $\epsilon_{61}=0$,
$\epsilon_{62}=0$, $\epsilon_{63}=R_1 B_0^2-R_2 B_0^4+R_3 B_0^6$, $\epsilon_{64}=\kappa_2$, $\epsilon_{65}=0$, $\epsilon_{66}=\sigma K -\Omega+R_1 B_0^2-R_2 B_0^4+R_3 B_0^6$.

The determinant of the system of algebraic equations must be equal to zero which results in the dispersion relation $\Omega=\Omega(K, q, p;
R_1, R_2, R_3, B_0)$ through which one can measure the instability gain spectra as $G({\Omega})=|\Im{\Omega_{max}}|$. Where, $\Im{\Omega_{max}}$ denotes the imaginary part of $\Omega_{max}$, where
$\Omega_{max}$ is the root of the polynomial with largest value.
\section{Modulational instability in metamaterial waveguide arrays}
\begin{figure*}[t]
	\subfigure[$R_1=1$ and $R_2=R_3=0$]{\label{3a}\includegraphics[width=0.32\linewidth]{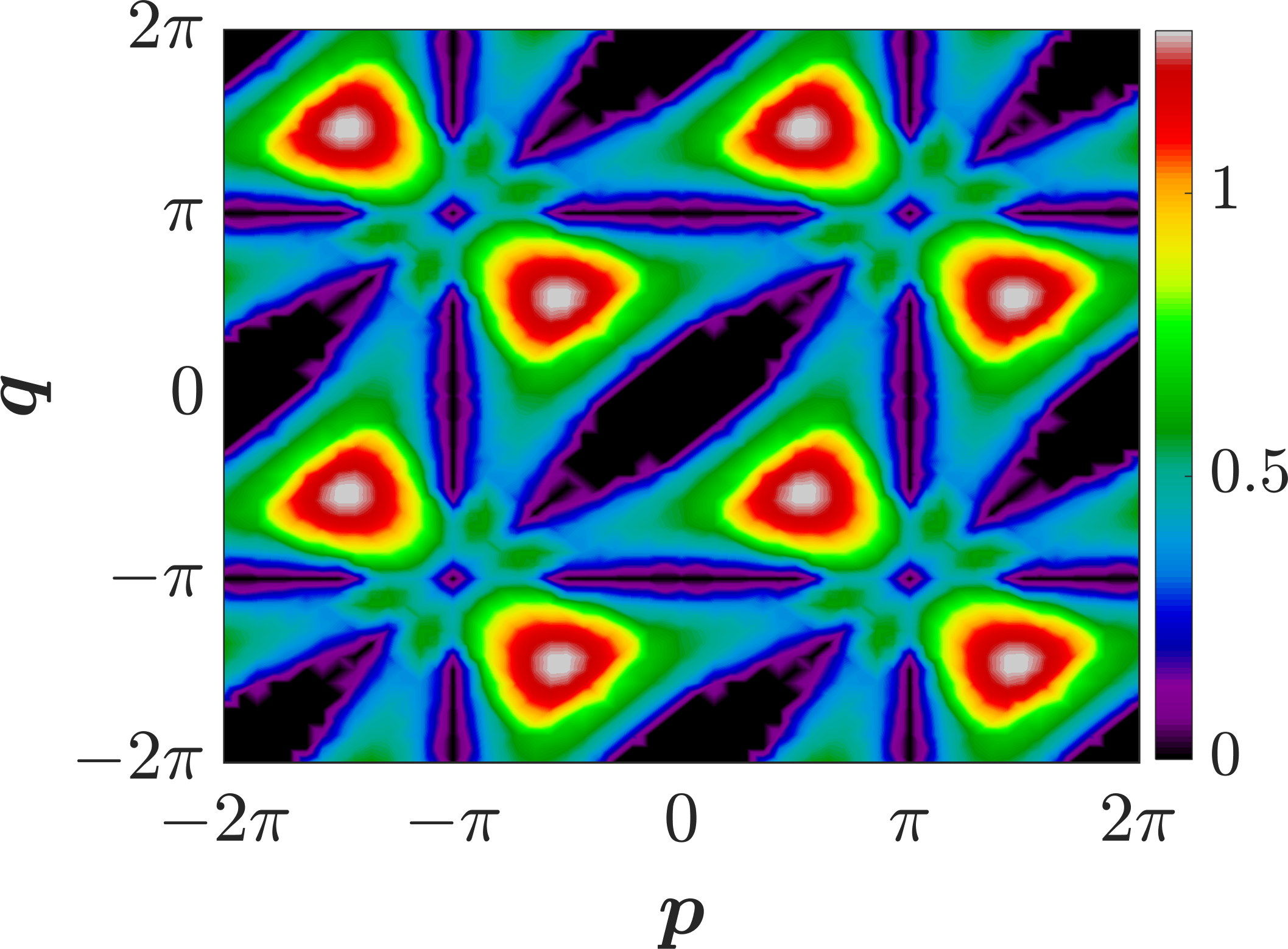}}
	\subfigure[$R_1=-R_2=1 $ and $R_3=0$]{\label{3b}\includegraphics[width=0.32\linewidth]{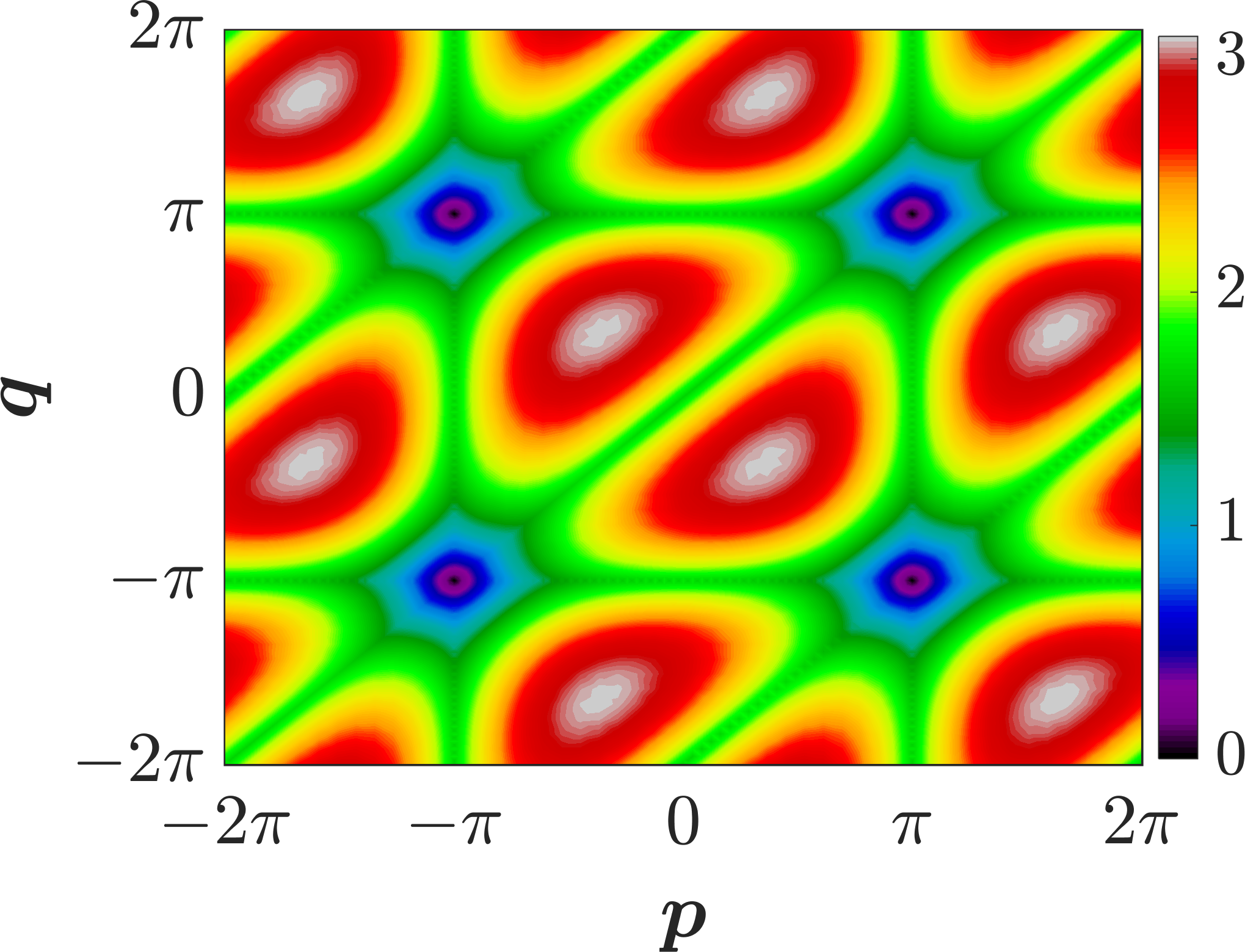}}\\
	\subfigure[$R_1=-R_2=R_3=1$]{\label{3c}\includegraphics[width=0.32\linewidth]{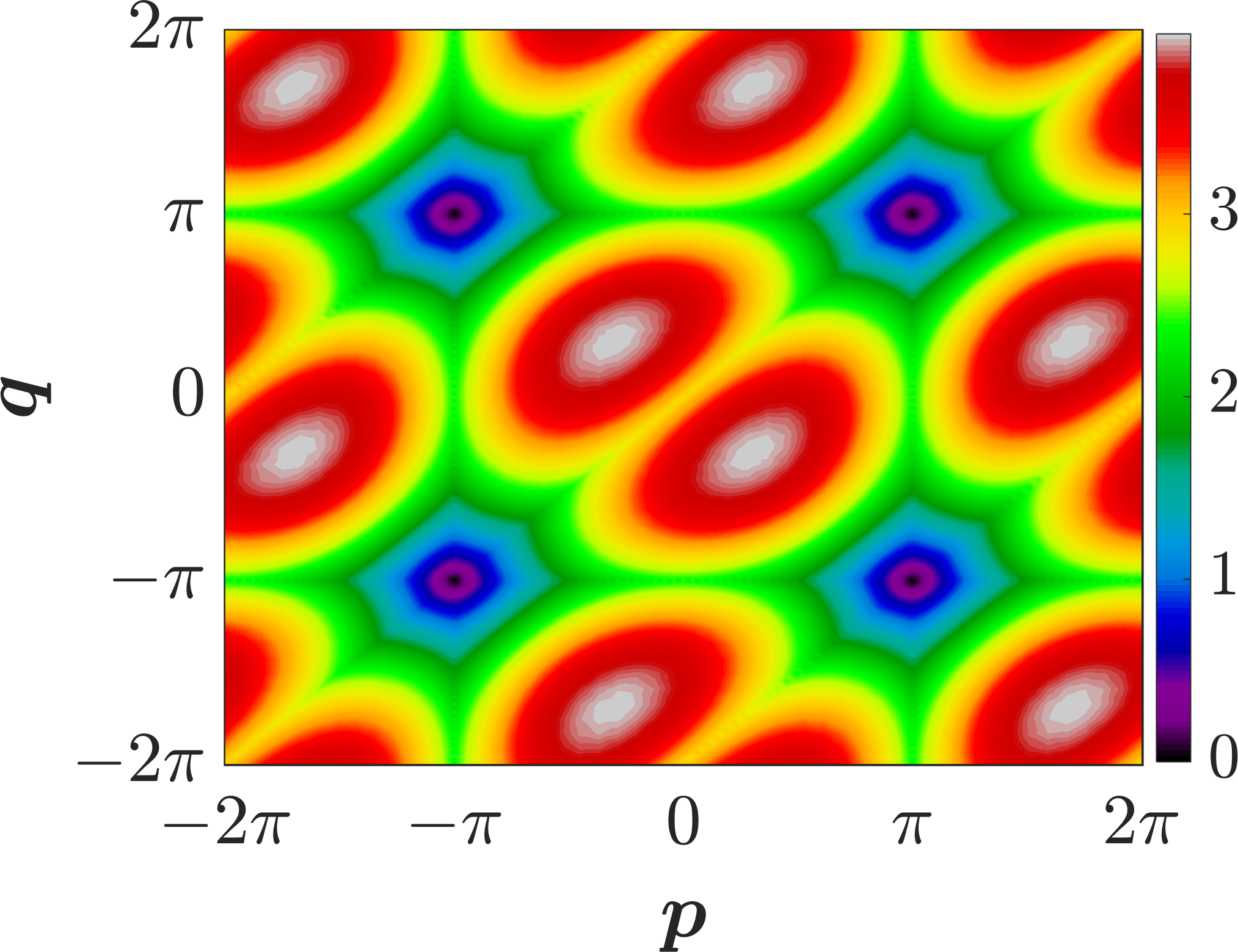}}
	\subfigure[$R_1=R_2=R_3=1$]{\label{3d}\includegraphics[width=0.32\linewidth]{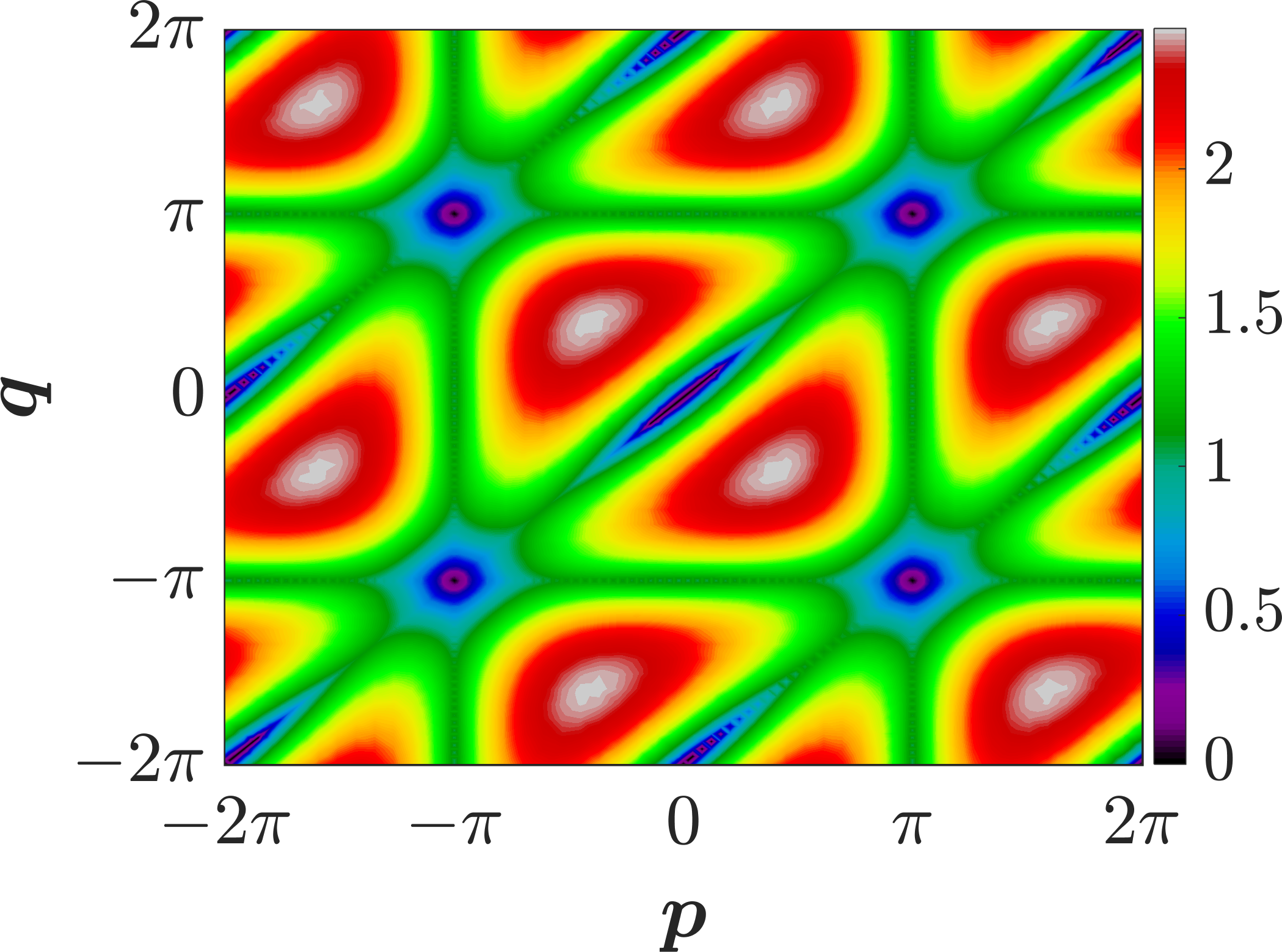}}
	\caption{(Color online.) Periodic MI gain spectra in the $p-q$ plane.}
	\label{3}
\end{figure*}
In this section we discuss the modulational instability of the flat band modes in the waveguide arrays with negative index material channels in detail. Let us choose initial power of incident wave, $P=B_0^2=1$. It is well known that higher order nonlinearities can considerably influence the system dynamics \cite{triki2016,triki2017,Raja1} and the modulation instability gain spectra in any system. Negative index materials embedded in cubic and quintic nonlinear media give more ways to manipulate and control modulation instability and hence the soliton formation \cite{sharma1}. Quintic nonlinearity plays a major role in the formation of gap solitons in fiber Bragg grating \cite{kp111}. The symmetric and asymmetric modulational instability growth rates have been observed in a zigzag array of nonlinear waveguides with the alternating signs of refractive indices \cite{ADD}. The self-focussing and self-defocussing nonlinearity0 of positive and negative refractive index waveguides affect the modulational instability gain of the array \cite{ADD1}. Modulational instability in the presence of  higher order nonlinearities may be beneficial to the generation of high repetition rate pulse trains in oppositely directed couplers \cite{malo}. In the same way, we investigate the influence of quintic and septimal nonlinearities on modulational instability in metamaterial waveguide arrays.
\par
It is well known that modulational instability is a precursor for the formation of solitons. The modulational instability of diffractionless modes can lead to bifurcations of the  modes to soliton-like solutions. Here we have given a special emphasis to analyze the influence of higher-nonlinear effects on modulational instability of diffractionless modes \cite{agp1, agp2}. Fig. \ref{MI1} depicts the instability gain spectra of diffractionless solution versus perturbation wave vector $K$ as a function of $p=q=S$ for different possible combinations of cubic, quintic and septimal nonlinearities.  Fig. \ref{MI1A}, which depicts the Kerr nonlinear case, it is clear from the figure that the MI spectra is periodic in $p=q=S$ with a period $2\pi$ (in the second Brillouin zone). In the Kerr nonlinear case each period consists of two instability regions, which are separated by a stable region located at $p=q=S=n\pi$ with $n=0,1,2,3...$.
\par
Now let us consider the influence of higher order nonlinearities originating from fifth and seventh order ( $\chi^{(5)}$ and $\chi^{(7)}$) susceptibilities. Fig. \ref{MI1B} represents the MI gain spectra with cubic, quintic and septimal nonlinearities. The quintic nonlinearity is of defocusing type whereas the cubic and septimal nonlinearities are of focusing types. Comparing with the cubic case (Fig. \ref{MI1A}), here one can see the enhancement of MI gain spectrum by increasing the gain and band width. In this case too, the periodic MI vanishes when the parameter $S$  satisfies the condition $S=n\pi$ with $n=0,1,2,3...$. Hence, the presence of non-Kerr nonlinearity enhances the modulational instability of diffractionless waves and provides more ways to manipulate solitons.

Fig. \ref{MI1C} depicts the case with focusing cubic and quintic nonlinearities alone. In this case also MI is periodic in $S$ with a period $2\pi$. It is
 interesting to note that compared to the previous cases (Figs. \ref{MI1A} and  \ref{MI1B}) here the MI gain is present for $S$ values of even integral multiples of $\pi$. The MI gain vanishes when $p=q=S=n\pi$ with $n=1,3,5...$. The presence of  focusing septimal nonlinearity enhances the MI by increasing the gain and enlarging the instability band as portrayed in Fig. \ref{MI1D}. The MI gain spectra for the case of focusing cubic and defocusing quintic and septimal nonlinearities is depicted in Fig. \ref{MI1E}. Here also one can understand the role of higher order nonlinearity in the enhancement of MI. Therefore, the stable propagation diffractionless mode and formation of localized soliton-like structure in non-Kerr photonic Lieb lattice with metamaterials can be achieved by properly tuning  higher-order nonlinearities.
\par
Fig. \ref{3} depicts the MI gain spectra in the $p-q$ plane for different combinations of nonlinearities. It is clear from Fig. \ref{3} that, the gain spectra are periodic in the $p-q$ plane with period $2\pi$. Stable propagation of diffractionless wave is observed when both $p$ and $q$ simultaneously satisfy the condition $p=q=n\pi$, $n=0,1,2,3...$ in Figs. \ref{3a} and \ref{3d}, where Fig. \ref{3a} corresponds to the cubic nonlinear case and in Fig. \ref{3d} cubic as well as septimal nonlinearities are of focusing types whereas quintic nonlinearity is of defocusing type. But the values of $n$ are odd numbers in Figs. \ref{3b} and \ref{3c}. It is either due to the presence of focusing quintic nonlinearity (Fig. \ref{3b}) or by the defocusing septimal nonlinearity Figs. \ref{3c}.
\par
One can thus conclude that the  stability of diffractionless waves propagating in a waveguide array with alternating signs of refractive index is highly influenced by the values of higher order nonlinear coefficients and the normalized coefficient functions $\kappa(k)$. Stable propagation of electromagnetic wave can be achieved by controlling these parameters in the lattice. Also, as the modulation instability is a precursor for the pattern formations in the form ultra-short pulses, controlling of modulation instability with these parameters provides better ways to generate and manipulate solitons in arrays of waveguides with negative index of refraction.
\section{Conclusion}
In summary, we have investigated the propagation of flat band modes in a face-centered square lattice of waveguide array, which is featured
by positive and negative refractive indices. We have considered three different waveguides with different
optical properties in a unit cell of the lattice. The study shows that the lattice supports dispersion curve with flat band and hence, it can bear diffractionless wave propagation. The photonic band structure as well as the stability of flat band modes are  highly dependent on coefficient functions $\kappa(k)$ and higher order nonlinearities.  Hence, the stable propagation of electromagnetic waves can be achieved by tuning these parameters in the lattice. Also, this study suggests the possibility for optically controlling the band structures of waveguide arrays.
We thus anticipate that the present investigation can pave a new roadmap on the nonlinear flat band modes which can be a promising candidate to control light by light in arrays of combined positive-negative index waveguides.
\section*{ Acknowledgement}
A. K. S. is very much grateful to Dr. K. Porsezian, who although is no longer with us but continues to inspire with his example and dedication, and who ignited the theme of the present work. The work of A.K.S. is supported by the University Grants Commission (UGC), Government of India, through a D. S. Kothari Post Doctoral Fellowship in Sciences.
The research of A.I.M. was supported by the Russian Foundation for
Basic Research (Grant No. 18-02-00278). A. G. is supported by the University Grants Commission (UGC), Government of India, through a D. S. Kothari Post Doctoral Fellowship in Sciences. M. L. is supported by a DST-SERB Distinguished Fellowship (Grant No. SB/DF/04/2017).
\begin{widetext}
\section*{Appendix}
In this Appendix, we provide the general solution to the dispersion equation (\ref{disp1}). It reads as

$$\omega_1=\frac{1}{3}N_1-[2^{1/3} (N_2+3 N_3)]/\{3 [N_4+\surd [N_4^2+4 (-N_2+3 N_3)^3]^{1/3}\}$$$$+\frac{1}{ 2^{1/3} 3}\{N_4
+\surd [N_4^2+4 (-N_2+3 N_3)^3]\}^{1/3},~~~~~~~~~~~~~~~~~~~~~~~~~~~~~~~~~~~~~~~~~~~~~~~~~~~~~~~~~~(A1)$$

$$\omega_2=\frac{1}{3}N_1+[(1+i \sqrt{3}) (N_2+3 N_3)]/\{ 2^{2/3} 3 [N_4+\surd [N_4^2+4 (-N_2+3 N_3)^3]^{1/3}\}$$$$-\frac{1}{6 2^{1/3}}(1-i \sqrt{3})\{ N_4
+\surd [N_4^2+4 (-N_2+3 N_3)^3]\}^{1/3},~~~~~~~~~~~~~~~~~~~~~~~~~~~~~~~~~~~~~~~~~~~~~~~(A2)$$

$$\omega_3=\frac{1}{3}N_1-[(1-i \sqrt{3}) (N_2+3 N_3)]/\{ 2^{2/3} 3 [N_4+\surd [N_4^2+4 (-N_2+3 N_3)^3]^{1/3}\}$$$$+\frac{1}{6 2^{1/3}}(1+i \sqrt{3})\{ N_4
+\surd [N_4^2+4 (-N_2+3 N_3)^3]\}^{1/3},~~~~~~~~~~~~~~~~~~~~~~~~~~~~~~~~~~~~~~~~~~~~~~(A3)$$

where
$$N_1=-f_1-f_2-f_3+k_z+2 \sigma  k_z,$$
$$N_2=(f_1+f_2+f_3-k_z-2 \sigma  k_z)^2,$$

$$N_3=-|\kappa_1|^2-|\kappa_2|^2+f_1 f_2+f_1 f_3+f_2 f_3-2 \sigma  f_1 k_z-f_2 k_z-\sigma  f_2 k_z-f_3 k_z-\sigma  f_3 k_z+2 \sigma  k_z^2+\sigma ^2 k_z^2$$

and
$$N_4=-9 |\kappa_1|^2 f_1-9 |\kappa_2|^2 f_1-2 f_1^3-9 |\kappa_1|^2 f_2+18 |\kappa_2|^2 f_2+3 f_1^2 f_2+3 f_1 f_2^2-2 f_2^3$$$$
+18 |\kappa_1|^2 f_3-9 |\kappa_2|^2 f_3+3 f_1^2 f_3-12 f_1 f_2 f_3+3 f_2^2 f_3+3 f_1 f_3^2+3 f_2 f_3^2-2 f_3^3+9 |\kappa_1|^2 k_z+9 |\kappa_1|^2 k_z$$$$
-9 |\kappa_1|^2 \sigma  k_z-9 |\kappa_2|^2 \sigma  k_z+6 f_1^2 k_z-6 \sigma  f_1^2 k_z-6 f_1 f_2 k_z+6 \sigma  f_1 f_2 k_z$$$$
-3 f_2^2 k_z+3 \sigma  f_2^2 k_z-6 f_1 f_3 k_z+6 \sigma  f_1 f_3 k_z+12 f_2 f_3 k_z-12 \sigma  f_2 f_3 k_z-3 f_3^2 k_z$$$$
+3 \sigma  f_3^2 k_z-6 f_1 k_z^2+12 \sigma  f_1 k_z^2-6 \sigma ^2 f_1 k_z^2+3 f_2 k_z^2-6 \sigma  f_2 k_z^2+3 \sigma ^2 f_2 k_z^2+3 f_3 k_z^2$$$$
-6 \sigma  f_3 k_z^2+3 \sigma ^2 f_3 k_z^2+2 k_z^3-6 \sigma  k_z^3+6 \sigma ^2 k_z^3-2 \sigma ^3 k_z^3.$$

It may be noted that the particular solutions given in Eq. (\ref{dis}) for the special case $f_2$ = $f_3$ follows from the above.
\end{widetext}


\begin{thebibliography}{100}
\bibitem{Flach:Leykam:14} S. Flach, D. Leykam, J.D. Bodyfelt, P. Matthies, and
A.S. Desyatnikov
\emph{Europhys. Lett.} \textbf{105}, 30001 (2014).

\bibitem{Longi:14} St. Longhi,
\textit{Opt. Lett.} \textbf{39}, 5892--5895 (2014).

\bibitem{Maimis:15} A. I. Maimistov, 
\textit{J.Phys. Conference Series} \textbf{613}, 012011 (2015)

\bibitem{G:Malomed:16} G. Gligoric, A. Maluckov, L. Hadzievski, S. Flach, and B. A.
Malomed,
\textit{Phys.Rev. B.} \textbf{94}, 144302 (2016) (8 pp).
\bibitem{Daniel}D. Leykam and S. Flach, APL Photonics \textbf{3}, 070901 (2018).
\bibitem{Leykam1}D. Leykam, A. Andreanov and S. Flach, Advances in Physics: X, \textbf{3}, 1473052 (2018).
\bibitem{Mukherjee:15} S. Mukherjee and R.R. Thomson.
\textit{Opt. Lett.} \textbf{40}(23), 5443-5446 (2015).


\bibitem{Mukherjee:Spracklen:15a} S. Mukherjee, A. Spracklen, D.
Choudhury, N. Goldman, P. Ohberg, E. Andersson, and R.R. Thomson,
\textit{Phys.Rev.Lett}. \textbf{114}, 245504 (2015).


\bibitem{Vicencio:14}
D. Guzman-Silva, C. Mejia-Cortes, M.A. Bandres, M.C. Rechtsman, S.
Weimann, S. Nolte, M. Segev, A. Szameit, and R.A. Vicencio,
\textit{New J. Phys}. \textbf{16}, 063061 (2014).

\bibitem{Vicencio:15} R.A.
Vicencio, C. Cantillano, L. Morales-Inostroza, B. Real, C.
Mejia-Cortes, St. Weimann, Al. Szameit, and M.I. Molina.
\emph{Phys.Rev.Lett.} \textbf{114}, 245503 (2015).

\bibitem{Fang:15} Y. T. Fang, H. Q. He, J. X. Hu, L. K. Chen, and Z.
Wen.
\textit{Phys. Rev. A.} \textbf{91}, 033827 (2015).

\bibitem{Maim:Gabi:16} A.I. Maimistov, I.R. Gabitov,
\textit{J.Phys. Conference Series} \textbf{714}, 012013 (4 pp)
(2016).


\bibitem{Maim:16} A.I. Maimistov,
\textit{Nonlinear Phenomena in Complex Systems}, \textbf{19},
358--367 (2016).
\bibitem{Mukherjee2}S. Mukherjee and R. R. Thomson,Opt. Lett.  \textbf{42}, 2243 (2017).
\bibitem{Bastian}B. Real, C. Cantillano, D. L. Gonzalez, A. Szameit, M. Aono, M. Naruse, S. J. Kim, K. Wang and R. A. Vicencio, Nature Scientific Reports \textbf{7} 15085 (2017).

\bibitem{ppbelicev}P. P. Belicev, G. Gligoric, A. Maluckov, M. Stepic and M. Johansson, Phys. Rev. A \textbf{96}, 063838 (2017).

\bibitem{Hamidreza}H. Ramezani, Phys. Rev. A \textbf{96}, 011802(R), 2017.
\bibitem{Shiqiang}S. Xia, Y. Hu, D. Song, Y. Zong, L. Tang, and Z. Chen, 
Opt. Lett. \textbf{41}, 1435-1438 (2016).
\bibitem{Yuanyuan}Y. Zong, S. Xia, L. Tang, D. Song, Y. Hu, Y. Pei, J. Su, Y. Li, and Z. Chen, 
     Opt. Express \textbf{24}, 8877-8885 (2016).
    \bibitem{vice}Rodrigo A. Vicencio and M. Johansson, Phys. Rev. A \textbf{87}, 061803(R) (2013).
    \bibitem{maimistov17} A. I Maimistov, J. Opt. \textbf{19} 045502 (2017).


\bibitem{Eisenberg}H. S. Eisenberg, Y. Silberberg, R. Morandotti, A. R. Boyd, and J. S. Aitchison, 
    Phys. Rev. Lett. \textbf{81}, 3383 (1998).
\bibitem{Morandotti}R. Morandotti, U. Peschel, J. S. Aitchison, H. S. Eisenberg, and Y. Silberberg, 
    Phys. Rev. Lett. \textbf{83}, 2726 (1999).
 \bibitem{Minardi}S. Minardi, F. Eilenberger, Y. V. Kartashov, A. Szameit, U. R\''{o}pke, J. Kobelke, K. Schuster, H. Bartelt, S. Nolte, L. Torner, F. Lederer, A. T\"{u}nnermann, and T. Pertsch, 
     Phys. Rev. Lett. \textbf{105}, 263901 (2010).
\bibitem{Falk}F. Eilenberger, S. Minardi, A. Szameit, U. Ropke, J. Kobelke, K. Schuster, H. Bartelt, S. Nolte, A. Tunnermann, and T. Pertsch, 
    Optics Express \textbf{19},  23171-231887 (2011).
\bibitem{Mandelik}D. Mandelik, R. Morandotti, J. S. Aitchison, and Y. Silberberg, 
Phys. Rev. Lett. \textbf{92}, 093904 (2004).
\bibitem{Shu}S. Jia, W. Wan, and J. W. Fleischer, 
Phys. Rev. Lett. \textbf{99}, 223901 (2007)
\bibitem{Zezyulin}D. A. Zezyulin, V. V. Konotop, and F. K. Abdullaev, 
Opt. Lett. \textbf{37}, 3930-3932 (2012).
\bibitem{Alexander}Alexander A. Dovgiy and Ilya S. Besedin, 
    Phys. Rev. E \textbf{92}, 032904 (2015).
\bibitem{Lingling}L. Zhang, Y. Xiang, X. Dai, and S. Wen, 
    J. Opt. Soc. Am. B  \textbf{31},  3029-3037 (2014).
\bibitem{shaf}K. Porsezian, A. K. Shafeeque Ali, and A. I. Maimistov, 
    J. Opt. Soc. Am. B \textbf{35}, 2057-2064 (2018).
 \bibitem{smith1}D. R. Smith, W. J. Padilla, D. C. Vier, S. C. Nemat-Nasser,
and S. Schultz, Phys. Rev. Lett. \textbf{84}, 4184
(2000).
\bibitem{shelby} R. A. Shelby, D. R. Smith, S. C. Nemat-Nasser, and S.
Schultz,  Appl. Phys. Lett. \textbf{78}, 489
(2001).
\bibitem{shelby2} R. A. Shelby, D. R. Smith, and S. Schultz, Science \textbf{292}, 77
(2001).
 \bibitem{scot} S. Townsend, D. S. Zhou, and Q. Li, Opt. Expr. \textbf{23}, 18236
(2015).
\bibitem{jason} J. Valentine, S. Zhang, T. Zentgraf, E. Ulin-Avila, D. A. Genov, G. Bartal and X. Zhang, Nature \textbf{455}, 376
(2008).

\bibitem{lapine}M. Lapine, M. Gorkunov, and K. Ringhofer,
Phys. Rev. E \textbf{67}, 065601 (2003).

\bibitem{agran} V. M. Agranovich, Y. R. Shen, R. H. Baughman, and A. A. Zakhidov, Phys. Rev. B \textbf{69}, 165112 (2004).
\bibitem{liang}L. Du, X. Zhou, and G. A. Fiete, Phys. Rev. B \textbf{95}, 035136 (2017).

\bibitem{Shumin}Shumin Xiao, V. P. Drachev, A.V. Kildishev, X. Ni, U. K.
Chettiar, H-K. Yuan, and V. M. Shalaev,  Nature \textbf{466}, 735 (2010).
\bibitem{Shad} I. V. Shadrivov, A. B. Kozyrev, D. W. van der Weide and Y. S. Kivshar,
Appl. Phys. Lett. \textbf{93}, 161903 (2008).

\bibitem{Tass} B. Dastmalchi, P. Tassin, T. Koschny and C. M. Soukoulis, Phys. Rev.
B \textbf{89}, 115123 (2014).


    \bibitem{triki2016}
    H.~Triki, K.~Porsezian, A.~Choudhuri, and P.~T. Dinda, 
    Phys. Rev. A \textbf{93}, 063810 (2016).

    \bibitem{triki2017}
    H.~Triki, K.~Porsezian, P.~T. Dinda, and P.~Grelu,
    Phys. Rev. A \textbf{95},
    023837 (2017).
    \bibitem{Raja1}
    S.~V. Raja, A.~Govindarajan, A.~Mahalingam, and M.~Lakshmanan,
    Phys. Rev. A \textbf{100}, 033838 (2019).

 \bibitem{sharma1} M. Saha and A. K. Sarma, Opt. Commun. \textbf{291}, 321 (2013).
   \bibitem{kp111}K. Porsezian, K. Senthilnathan, and S. Devipriya, J. Quan. Electronics, \textbf{41}, 789 (2005).
  \bibitem{ADD} A. A. Dovgiy, Quantum Elec. \textbf{44}, 1119, (2014).
   \bibitem{ADD1}L. Zhang, Y. Xiang, X. Dai and S. Wen, J. Opt. Soc. Am. B \textbf{31}, 3029 (2014).
  \bibitem{malo} A. Mohamadou,P.H. Tatsing, C.G. Latchio Tiofack, C.B. Tabi  and T.C. Kofane, J.Modern Opt. \textbf{61}, 1670 (2014).
   \bibitem{agp1} G. P. Agrawal, Nonlinear Fiber Optics (4th edition). Academic Press, San Diego (2007). 
    \bibitem{agp2} G. P. Agrawal, Application of Nonlinear Fibre Optics (5th edition). Academic Press, San Diego (2012).
\end{thebibliography}
\end{document}